\documentclass[smallextended]{svjour3}
\pdfoutput=1
\date{2023-05-18. Preprint.\newline Accepted version available at: \href{http://dx.doi.org/10.1007/s10664-023-10377-w}{http://dx.doi.org/10.1007/s10664-023-10377-w}}
\usepackage{algorithmic}
\usepackage{amsmath,amssymb,amsfonts}
\usepackage{cite}
\usepackage{comment}
\usepackage[l3]{csvsimple}
\usepackage{enumerate}
\usepackage{fancyvrb}
\usepackage{graphicx}
\usepackage{lipsum}
\usepackage{listings}
\usepackage[colorlinks,urlcolor=black,linkcolor=black,citecolor=black,linktocpage=true]{hyperref}
\usepackage{pifont}
\usepackage{siunitx}
\usepackage{subfigure}
\usepackage{textcomp}
\usepackage{todonotes}
\usepackage{xcolor}
\usepackage{xspace}

\lstdefinestyle{myPython}{
  language=Python,
  morekeywords={with,as},
  basicstyle=\ttfamily,
  showstringspaces=false,
  commentstyle=\itshape\color{gray},
  keywordstyle=\color{teal},
  stringstyle=\color{cyan},
  basicstyle=\ttfamily\small,
}

\newcommand{\OK}{\mbox{\ding{51}}}
\newcommand{\KO}{\mbox{\ding{55}}}

\newcommand{\SWH}{Software Heritage\xspace}
\newcommand{\swhid}[1]{\href{https://archive.softwareheritage.org/#1}{{\small\ttfamily #1}}}

\newcommand{\revchange}[1]{#1}  

\def\DataBlobCountApproxM{6.9}
\def\DataBlobCountApprox{\DataBlobCountApproxM\,M}
\def\DataEdILabel{2019-03-21}
\def\DataEdIILabel{2021-03-23}
\def\DataEdIIILabel{2022-04-25}
\def\DataBlobCountApproxEdI{3.4\,M}
\def\DataBlobCountApproxEdII{6.5\,M}
\def\DataBlobCountApproxEdIII{\DataBlobCountApprox}
\def\DataBlobCount{6859189}

\def\DataBlobNoOriginCount{1254}
\def\DataBlobNoOriginPct{0.02}
\def\DataTarSizeApprox{13\,GiB}
\def\DataTarExpandedSizeApprox{84\,GiB}
\def\DataMimeTextPlainPct{84}
\def\DataMimeTextPct{98}
\def\DataSwhOriginCountApproxM{186}
\def\DataGplDifferentNames{662}

\def\DocumentsManualCount{8102}
\def\DocumentsManualNormalizedCount{4371}

\begin{document}
\title{The Software Heritage License Dataset (2022 Edition)}
\author{Jesus M.~Gonzalez-Barahona\and
  Sergio Montes-Leon\and
  Gregorio Robles\and
  Stefano Zacchiroli}

\institute{
  J.~M.~Gonzalez-Barahona
  \at Universidad Rey Juan Carlos, Madrid, Spain\newline
  \email{jesus.gonzalez.barahona@urjc.es}
  \and
  S.~Montes-Leon
  \at Universidad Rey Juan Carlos, Madrid, Spain\newline
  \email{s.montesl@alumnos.urjc.es}
  \and
  G.~Robles
  \at Universidad Rey Juan Carlos, Madrid, Spain\newline
  \email{gregorio.robles@urjc.es}
  \and
  S.~Zacchiroli
  \at LTCI, Télécom Paris, Institut Polytechnique de Paris, Paris, France\newline
  \email{stefano.zacchiroli@telecom-paris.fr}
}

 \maketitle
\begin{abstract}

  \emph{Context:} When software is released publicly, it is common to include with it either the full text of the license or licenses under which it is published, or a detailed reference to them. Therefore public licenses, including FOSS (free, open source software) licenses, are usually publicly available in source code repositories.
  
  \emph{Objective:} \revchange{To compile a dataset containing as many documents as possible} that contain the text of software licenses, or references to the license terms. Once compiled, characterize the dataset so that it can be used for further research, or practical purposes \revchange{related to license analysis}.

  \emph{Method:} Retrieve from Software Heritage---the largest publicly available archive of FOSS source code---all versions of all files whose names are commonly used to convey licensing terms. All \revchange{retrieved} documents will be characterized in various ways, using automated and manual analyses.

  \emph{Results:} The dataset consists of \DataBlobCountApproxM{} million unique license files. Additional metadata about shipped license files is also provided, making the dataset ready to use in various contexts, including: file length measures, MIME type, SPDX license (\revchange{detected} using ScanCode), and oldest appearance. The results of a manual analysis of \DocumentsManualCount{} documents \revchange{is also included, providing} a ground truth for further analysis. The dataset is released as open data as an archive file containing all deduplicated license files, plus several portable CSV files with metadata, referencing files via cryptographic checksums.

  \emph{Conclusions:} \revchange{Thanks to the extensive coverage of} Software Heritage, the dataset presented in this paper covers a very large fraction of all software licenses for public code. We have assembled a large body of software licenses, \revchange{characterized it quantitatively and qualitatively}, and validated that it is mostly composed of licensing information and includes almost all known license texts. The dataset can be used to conduct empirical studies on open source licensing, training of automated license classifiers, natural language processing (NLP) analyses of legal texts, as well as historical and phylogenetic studies on FOSS licensing. It can also be used in practice to improve tools detecting licenses in source code.

\end{abstract}

\keywords{dataset, open source, software license, copyright, intellectual
  property, software engineering, natural language processing}

\section{Introduction}
\label{sec:intro}

Many different software licenses exist and are used in public code. Some of them are considered as proper ``open source'' \revchange{licenses} by OSI (Open Source Initiative). Others (with a significant overlap) are labeled as ``free software'' by the FSF (Free Software Foundation). Still some others are neither ``open source'' nor ``free'' but are applied to software components distributed in source code form, for example via GitHub or GitLab. In any case, licenses should be considered, and respected, when reusing those components. This is the reason why identifying the license or licenses of a software component is so important for those reusing or extending it.

Each license \revchange{comes with its} own licensing terms~\cite{gomulkiewicz2009open}. They are so varied that industry standards like SPDX (Software Package Data Exchange) emerged to normalize license naming and identifiers~\cite{stewart2010spdxspec}. The SPDX work group of the Linux Foundation also published ``License Inclusion Principles'',\footnote{License Inclusion Principles: \\
  \url{https://github.com/spdx/license-list-XML/blob/main/DOCS/license-inclusion-principles.md}} which we use as a definition of \revchange{what a license file is}: the document used by the copyright owner to detail the permission given to those who receive the software. In the case of publicly available software, including FOSS (free, open source software), the license is usually included as a file with the distributed source code. It is this license which gives, in the case of FOSS, permission to reuse, redistribute, and extend it, subject to certain conditions that vary from license to license~\cite{rosen2005osslicensing, lindberg2008osslicensing}.

Proper management of an increasingly complex software supply chain~\cite{harutyunyan2020osssupplychain} requires being able to deal with many license combinations, their potential incompatibility~\cite{german2012liccompatibility}, and auditing increasingly large code bases, ideally in an automated way~\cite{phipps2020continuouscompliance}. These real-world needs have motivated over the years several empirical software engineering studies on the evolution of open source
licensing~\cite{manabe2010licensesevol, debsources-ese-2016,
  vendome2017githublicenses}, on the emergence of open source \emph{license
  variants} and exceptions~\cite{german2015licensevariability,
  vendome2017licexceptions}, as well as the development of industry-strength
tools to automatically detect and classify (FOSS)
licenses~\cite{gobeille2008fossology, german2015licensevariability,
  ombredanne2020sca}.

In this context, the main objective of the efforts reported in this paper is to produce a dataset that helps to better understand licenses used in publicly available source code. We intend such dataset to include most, if not all, license texts used when publishing publicly accessible software. The documents in the dataset are augmented with automatically obtained metadata, and with the manual annotation of a sample of them, with the means of making it easy to conduct further analyses. As a result, we contribute with: (i) the largest open dataset of license texts and related information, (ii) a detailed description and characterization of the dataset, (iii) examples of use and auxiliary information and tools for making it easier working with the dataset, and (iv) some \revchange{preliminary} results and findings obtained by analyzing the dataset.

\subsection{The dataset}

The dataset presented in this paper\footnote{\revchange{The version of the dataset discussed in this paper is available at \url{https://annex.softwareheritage.org/public/dataset/license-blobs/2022-04-25/}; other versions of the dataset (both past versions and future ones) are available starting from \url{https://annex.softwareheritage.org/public/dataset/license-blobs/}}} is composed of the following elements:

\begin{enumerate}
\item The \textbf{document collection}, which includes \num{\DataBlobCount} unique documents obtained from \SWH, the largest public archive of software source code, by querying for all filenames that likely \revchange{contain} licensing information. \SWH assembled the largest collection of publicly available software source code, with a total of \DataSwhOriginCountApproxM{} million public software origins including public Git repositories (from GitHub and GitLab), FOSS distributions (e.g., Debian), and package manager repositories (e.g., PyPI, NPM).\footnotemark{}

\footnotetext{\revchange{\SWH is an archival project established in 2015 with the stated goal of: collect, preserve forever, and make publicly available the entire body of software, in the preferred form for making modifications to it. A detailed description of the project if out-of-scope for this paper, therefore we refer the interested reader to: previous publications about the project~\cite{swhipres2017, swhcacm2018}, its homepage at \url{https://www.softwareheritage.org}, and the archive status page at \url{https://archive.softwareheritage.org} (accessed 2022-10-20) where one can find an up-to-date view of the software origins that are periodically crawled to populate the archive.}}
  
\item \textbf{Metadata} about all \revchange{license documents in the collections}: file names, length measures, detected MIME type, contained FOSS licenses detected using ScanCode~\cite{scancode-toolkit}, example origin, oldest and total number of public commits in which the license file appears.

\item An \textbf{annotated sample} of \DocumentsManualCount{} manually reviewed documents, randomly sampled from the whole dataset. Documents in this sample has checked to assess if they include license text, a reference to one or multiple licenses, a copyright statement and other details (e.g., if the license was correctly identified by ScanCode). The annotated sample is suitable for being used as a \textbf{ground truth} for further studies.

\end{enumerate}

\subsection{Analysis of the dataset}

\revchange{As a ``starter kit'' for using the dataset} we analyze it to answer the following research questions:

\begin{itemize}
\item \textbf{RQ1}: How many distinct licenses \revchange{does the dataset contain}?

  This question intends to measure the diversity of licenses in the dataset. As any change to a license text may result in the terms and conditions being changed, it is relevant to know if these changes happen frequently or not.

  To answer this question, we will identify the number of distinct files containing a single license, therefore considering only license files with the full text of a license. The resulting number should be the minimum number of license texts that any license-text detection tool should recognize. On the other hand, this will allow to estimate the size of the subset of files containing single full-text licenses in the dataset. \revchange{We will avoid files with more than one license, so that we are sure not to double count license texts that could be repeated, but accompanied by different licenses in different files.}
  
  Our results show that the total number of files containing a license (i.e., distinct license files that contain the full text of a license) is in the order of millions.
  
\item \textbf{RQ2}: Why are there so many distinct licenses in the \revchange{dataset}?

  This question is \revchange{relevant} because the number of distinct licenses usually considered in the state-of-the-art is relatively small. For example, OSI\footnote{OSI (Open Source Initiative): \url{https://opensource.org}} recognizes 68 open source licenses,\footnote{OSI Approved licenses: \url{https://opensource.org/licenses-draft} (accessed on 2022-10-30)} although they list a total of 96 licenses when including those that are considered superseded or retired. The more lax SPDX license list\footnote{SPDX license list: \url{https://spdx.org/licenses/} (accessed on 2022-10-30)} includes the text of 498 licenses. The largest public software license list is, to our knowledge, the ScanCode LicenseDB,\footnote{ScanCode LicenseDB: \\ \url{https://scancode-licensedb.aboutcode.org/} (accessed on 2022-10-30)} which currently contains the text of 1879 licenses.

  To answer this question, we will use several approaches: manual inspection of the ground truth, normalization of all license texts in the dataset, and automatic detection with ScanCode.

  As a result, we offer evidence that single-line copyright notices, blanks and upper/lowercase are a very important cause of the diversity of documents having full-text licenses.

\end{itemize}

\subsection{Data availability}

The dataset~\cite{jesus_m_gonzalez_barahona_2023_8200352} is released as open data, together with a replication package to recreate it from scratch.
It is available for download from Zenodo at \url{https://doi.org/10.5281/zenodo.8200352}.
The dataset consists of a \texttt{tar} archive containing unique license blobs (deduplicated based on SHA1 checksums) in a shared directory structure, together with a set of portable CSV and JSON files with derived metadata and for the manually annotated subset, all cross-referenced to license blobs via SHA1 checksums.

This is the third edition of the dataset, labeled \DataEdIIILabel.
The first edition, labeled \DataEdILabel, was distributed informally.\footnote{\url{https://annex.softwareheritage.org/public/dataset/license-blobs/2019-03-21/} (accessed 2022-11-10)}
The second edition, labeled \DataEdIILabel, was presented at the 2022 Mining Software Repositories Conference (MSR 2022)~\cite{msr-2022-foss-licenses},\footnote{\url{https://annex.softwareheritage.org/public/dataset/license-blobs/2021-03-23/} (accessed 2022-11-10)} and produced using data available in \SWH up to March 2021.
This third edition\footnote{\url{https://annex.softwareheritage.org/public/dataset/license-blobs/2022-04-25/} (accessed 2022-11-10)} was produced with data available in \SWH up to April 2022 and includes both a larger corpus and richer metadata with respect to previous editions. New releases of the dataset will be periodically released in the future.\footnote{\revchange{All dataset versions are available starting from \url{https://annex.softwareheritage.org/public/dataset/license-blobs/}}}

\subsection{Structure of this paper}

The remainder of this paper is structured as follows.
In the next section we offer a detailed description of the dataset, including the document collection, the metadata files and the annotated sample. Section~\ref{sec:methodology} outlines the method used to select, retrieve, and annotate (both automatically and manually) the dataset. A characterization of the dataset can be found in Section~\ref{sec:characterization}, with information on the main parameters, the licenses found, and a description of the annotated sample. Section~\ref{sec:findings} reports on the answers to the research questions and some other findings. Section~\ref{sec:usage} shows some hands-on example of dataset usage. After discussion (Section~\ref{sec:discussion}) and threats to validity (Section~\ref{sec:threats}), we present related work in Section~\ref{sec:related}. Conclusions are drawn in Section~\ref{sec:conclusion}.

 \section{Description of the dataset}
\label{sec:description}

The dataset presented in this paper is composed of three parts:

\begin{enumerate}
\item \textbf{The Document Collection}. This is a collection of documents obtained from \SWH. This collection is composed of all versions of all files in \SWH with names normally used for indicating the license of some source code (see details in Section~\ref{sec:methodology}). The documents collection includes many different types of documents: most of them are related to software licensing, but there are also other kinds. In the case of documents related to software licensing, they may include the whole text of a single license, but also just a notice with a reference to the actual license.\footnote{See the ``How to apply the Apache License to your work'' part of the Apache 2.0 license for an example of a license reference: \url{https://www.apache.org/licenses/LICENSE-2.0} (accessed 2022-11-10).} Documents may also contain, in the case of software compilations, or software including files from different projects, a list of licenses and license notices. Although in most cases documents are plain text files, other formats occur in the dataset as well, like PDF and HTML.

\item \textbf{The Metadata Files}. To facilitate the analysis of the Document Collection, we include extensive metadata about them. These Metadata Files provide information about every single document in the collection. Some information is about document characteristics (such as format or length), and other is specifically licensing information (as a result of running a tool for identifying licenses on all of them). These Metadata Files might be usable by themselves for several kinds of studies: they allow for a quick characterization of relevant aspects of documents in the collection.

\item \textbf{The Annotated Sample}. To get more insight about documents in the collection, we have manually inspected a large random sample of them. As a result, we have produced a ground truth of what documents in the collection contain actual software licenses, and shipped it as part of the dataset.
\end{enumerate}

\noindent
Although several editions of the dataset were produced over time (at the time of writing: 2019, 2021, and 2022, according to the date of collection), in this section we describe only the 2022 edition, as it is a superset of previous ones, due to the accumulative nature of \SWH. Only when relevant for comparison purposes, we present data for the other two editions. The Annotated Sample is novel and only shipped as part of the 2022 edition of the dataset.

\subsection{Document collection}
\label{subsec:description-collection}

The Document Collection is composed of \num{\DataBlobCount} documents, shipped in a single \texttt{tar} archive file (\texttt{blobs.tar.zst}) compressed with Zstandard~\cite{zstandard-rfc} and weighting approximately \DataTarSizeApprox. Each document is a ``\emph{blob}'' (a unique sequence of bytes) corresponding to one version of one file whose filename was included in the list of filenames initially retrieved from \SWH. The same blob may occur in different versions of different projects and under different filenames, if all of their contents are byte-by-byte identical. For example, if a license is included in different repositories, in exactly the same binary form but with different names, all of these files will be represented in the Document Collection by a single blob. On the contrary, if a license is present in different binary forms, even if they differ only in a single byte (an added newline, for example) each of those files will be a different blob.

Documents (blobs) are organized in the \texttt{tar} archive in a two-level-deep shared directory structure, based on the SHA1 checksum of each file. The filename used for each document is its SHA1 checksum serialized in hexadecimal form. The first-level directory names are composed of the first two characters of all the filenames in each of them, and the second-level directory names are composed of the second two characters of all filenames in each of them. For example, the path of the following document in the \revchange{expanded} archive: \texttt{blobs/02/52/0252d93ad297ec183a567ee813ab8c8d61ece655} corresponds to a license document whose SHA1 checksum is\\ \texttt{0252d93ad297ec183a567ee813ab8c8d61ece655}. Documents are \revchange{hence} fully deduplicated in the dataset based on SHA1 checksums: each document (blob) will appear only once in the collection.

For convenience, the dataset also includes \texttt{blobs-sample20k.tar.zst}, a smaller archive containing only \num{20 000} randomly selected license files. This smaller dataset can be used to perform a fast inspection and conduct trial experiments on a small scale before attacking the entire corpus.

\subsection{Metadata files}
\label{sec:datamodel}

\begin{figure}
  \center
  \includegraphics[width=\textwidth]{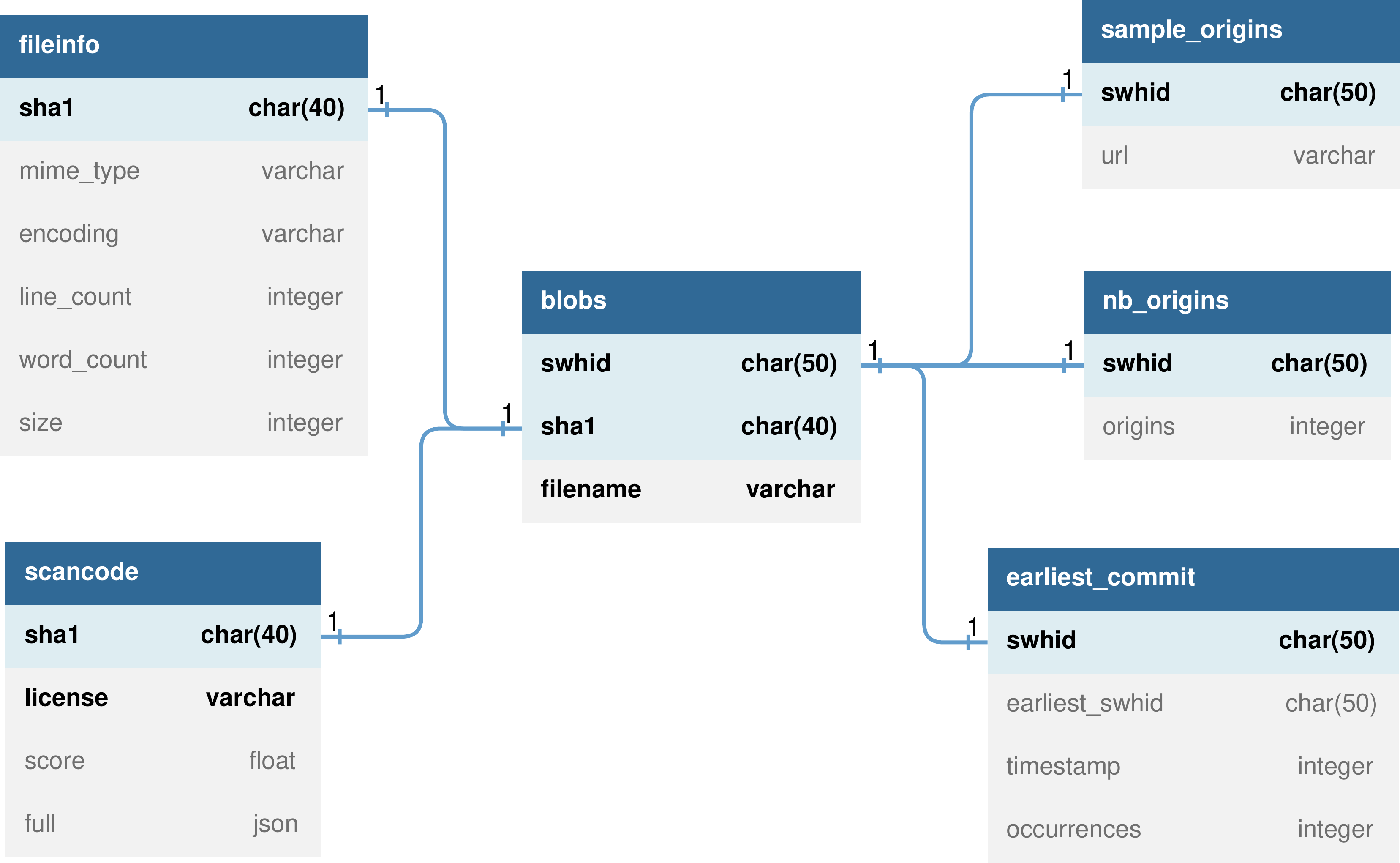}
  \caption{Relational data model for metadata files.}
  \label{fig:metadata-schema}
\end{figure}

Metadata for all documents in the collection are provided as a set of textual CSV~\cite{csv-rfc} and JSON files, compressed with Zstandard. Each CSV file corresponds to a table in the relational model shown in Figure~\ref{fig:metadata-schema}. \revchange{CSV files} can be used as such (for example, imported in R or \revchange{Pandas} data frames), or easily imported into an actual database management system. Metadata can be cross-referenced to the actual documents (in \texttt{blobs.tar.zst}) using \revchange{blob} checksums as keys.

Each table/CSV file captures the metadata described below:
\begin{itemize}
\item \texttt{\bfseries blobs} (CSV file: \texttt{license-blobs.csv.zst}) is the
  master index of all documents/blobs in the dataset. Each row in the file corresponds to a filename for a given document\footnote{If the document was found under several different filenames, as it could happen, it will appear in the index once for each different filename} and contains three columns:
  \begin{itemize}
  \item \texttt{sha1}: the SHA1 checksum of the document (blob).
  \item \texttt{swhid}: the Software Heritage persistent identifier (SWHID)~\cite{swhipres2018} of the document \revchange{(blob)}.
    SWHIDs are standard, persistent, intrinsic identifiers that can reference various kinds of software artifacts (files, directories, commits, releases, etc.) commonly found in Version Control Systems (VCS).
    For example, the SWHID for \revchange{the file containing} a popular form of the GPL version 3 text is:
      \swhid{swh:1:cnt:94a9ed024d3859793618152ea559a168bbcbb5e2}.
    \revchange{Similarly to plain SHA1 checksums, SWHIDs are computed by applying a cryptographic checksum function (currently: SHA1) to the digital manifestation, or content, of software artifacts.
    SWHIDs are explicitly typed, versioned, and more expressive than bare SHA1s.
    SWHIDs version 1 (used in this dataset) are also} compatible with the \revchange{object} identifiers used by the popular Git VCS, \revchange{whereas SHA1s are not, due to the ``salting'' added by Git before computing SHA1s}.

    In the context of this dataset both SWHIDs and SHA1 are used as keys for license documents, depending on the tables.
    Hence the main dataset index contains \emph{both} identifiers for each blob and can be used as a translation table between the two.
  \item \texttt{filename}: the \texttt{filename} given to a given document in a given context (e.g., one or more commits in a public Git repository). For example, the aforementioned variant of the GPL version 3 text is found with \DataGplDifferentNames{} different names, including \texttt{"COPYING"}, \texttt{"LICENSE.GPL3"}, and \texttt{"a2ps.license"}, which means there will be \DataGplDifferentNames{} lines in this table for that blob.
  \end{itemize}

  Both \texttt{swhid} and \texttt{sha1} are used by other tables as foreign key targets. There is no unique primary key column in this table, due to multiple filenames associated to each document.

\item \texttt{\bfseries fileinfo} (CSV file: \texttt{blobs-fileinfo.csv.zst}) provides basic information and size measures about documents. The main columns in this file are:
  \begin{itemize}
  \item \texttt{sha1}: document identifier, cross-reference to the \texttt{blobs} file.
  \item \texttt{mime\_type}, \texttt{encoding}: document MIME type and character encoding, as detected by \texttt{libmagic}~\cite{file-opengroup}.
  \item \texttt{size}: document size in bytes.
  \item \texttt{line\_count}, \texttt{word\_count}: report file sizes in lines and (blank-separated) words, respectively, for textual files.
  \end{itemize}

\item \texttt{\bfseries scancode} (CSV file: \texttt{blobs-scancode.csv.zst} and NDJSON file\linebreak \texttt{blobs-scancode.ndjson.zst}) reports about the license(s) contained in a given document, as detected by the ScanCode toolkit~\cite{scancode-toolkit, ombredanne2020sca}.\footnote{\revchange{Version used: \texttt{ScanCode 31.2.1}.}} Multiple license texts, or notices of licenses, can be detected within a single document, due to either multiple license texts being included or to different confidence levels as reported by ScanCode. The main columns in this file are:
  \begin{itemize}
  \item \texttt{sha1}: document identifier, cross-reference to \texttt{blobs} file.
  \item \texttt{license}: license found (either in full text, or as a license notice) in the document, expressed using the SPDX industry-standard~\cite{stewart2010spdxspec,gandhi2018spdx} identifier (e.g., \texttt{"GPL-3.0-only"}).
  \item \texttt{score}: ScanCode confidence level as a float in the $[0,100]$ range (100 being maximum confidence).
  \item \texttt{full}: ``virtual'' column containing the complete ScanCode results for each document in the dataset.
    This column is virtual in the sense that it is not actually present in the CSV file, but rather materialized in the Newline Delimited JSON file \texttt{blobs-scancode.ndjson.zst}.
    The file contains one JSON-document per line, where each JSON file contains the complete ScanCode results for one license document in the dataset.
    The JSON document is a dictionary with three keys:
    \begin{itemize}
    \item \texttt{sha1}: document SHA1.
    \item \texttt{licenses}: complete output of \texttt{scancode --license}, i.e., complete information about the licenses detected by ScanCode.\footnote{Details about the JSON schema: \\ \url{https://scancode-toolkit.readthedocs.io/en/stable/cli-reference/output-format.html} (accessed 2022-11-09)}
    \item \texttt{copyrights}: complete output of \texttt{scancode --copyright}, i.e., complete information about the copyright notices detected by ScanCode.
    \end{itemize}
  \end{itemize}
  
\item \texttt{\bfseries sample\_origins} (CSV file: \texttt{blobs-origins.csv.zst}) contains information about where documents were found, i.e., which repositories (or, more generally, ``software origins'' as they could be also packages in package repositories) have distributed them in the past. As each document can be distributed by tens of millions of repositories, only a single example of an origin is given for each document (\texttt{url} column). Obtaining from \SWH a list of \emph{all} repositories known to ship a given document is possible~\cite{swh-provenance-emse}, but out of scope for this dataset. For example, the aforementioned variant of the GPL-3 text was found (among others) in the Git repository at \url{https://github.com/tizenorg/platform.upstream.qtbase}.

  The file contains two columns:
  \begin{itemize}
  \item \texttt{swhid}: the document SWHID
  \item \texttt{url}: URL of \emph{a} sample origin where the document has been observed
  \end{itemize}

\item \texttt{\bfseries nb\_origins} (CSV file: \texttt{blobs-nb-origins.csv.zst}) contains rough popularity information about license documents, measured as the number of origins that have distributed them in the past.
  The file contains two columns:
  \begin{itemize}
  \item \texttt{swhid}: the document SWHID
  \item \texttt{origins}: number of origins observed as having distributed the license document in the past
  \end{itemize}
  For instance, the specific version of the GPL3 license text discussed above has been observed in \num{2 822 260} \revchange{\emph{different}} software origins. (\revchange{We recall from before that a software origin stands for a source code distribution place, e.g., a Git repository located at a given URL or a package in a package manager repository. In this example, about 2.8\,M such places distribute or have distributed in the past a specific version of the GPL3 license text}).

\item \texttt{\bfseries earliest\_commit} (CSV file: \texttt{blobs-earliest.csv.zst}) provides historical and (additional) popularity information:
  \begin{itemize}
  \item \texttt{swhid}: the document SWHID
  \item \texttt{earliest\_swhid}: SWHID of the oldest known public commit that contained the license file.
    In this table SWHIDs are used to reference both license documents, in the \texttt{swhid} column, and commits, in this column. The two can be distinguished by a) the position of the column, and b) due to the fact that SWHIDs are typed with, respectively, explicit \texttt{cnt} and \texttt{rev} strings.
    For example, the following commit contains a variant of the MIT license that includes a Russian copyright notice:
    \swhid{swh:1:rev:088313246501c78ae9d7f08e46aaea45855c5c7e}.
    The referenced commit can be then looked up using the Software Heritage Web UI, Web API, or locally on the filesystem using swh-fuse~\cite{swh-fuse}.
    For example, said Russian MIT variant can be browsed at
    {\small \url{https://archive.softwareheritage.org/swh:1:rev:088313246501c78ae9d7f08e46aaea45855c5c7e}} \linebreak (accessed 2022-10-30).

  \item \texttt{timestamp}: the commit timestamp, as Unix time.
  \item \texttt{occurrences}: the total number of commits known by Software Heritage to contain the document. It can be used as another rough measure of document popularity---in addition to the number of origins from the \texttt{nb\_origins} table.
  \end{itemize}

\end{itemize}

\subsection{Annotated sample}
\label{subsec:description-sample}

We have also manually analyzed a subset of \DocumentsManualCount{} documents chosen randomly from the whole collection. We have annotated them with several characteristics that help to understand what kind of documents the dataset contains, but also to conduct further studies. \revchange{The obtained annotations are shipped as a set of CSV files:}
\begin{itemize}

\item \texttt{\bfseries truth} (CSV file: \texttt{truth.csv}) is a file that can be used as ground truth, for comparing results by any study identifying files with a single full license text, or with a single license notice, etc. Its main columns are:
  \begin{itemize}
  \item \texttt{name}: \revchange{The document identifier, as a SHA1 checksum.}

  \item \texttt{uhash}: SHA1 hash of the normalized text of the document. All plain text documents in the sample were normalized (removing blanks, lowercasing, and removing one-line copyright notices), and SHA1 was computed on the resulting text, to ease detection of documents with only ``cosmetic'' differences.

  \item \texttt{length}, \texttt{codec}, \texttt{mime}, \texttt{scancode}: Data corresponding to the document in Metadata Files, to ease cross-analysis with these fields.

  \item \texttt{licen} (Boolean): True if the file includes one and only one license text, and possibly something else, not related to licensing (in which case, \texttt{selse} will also be true). False otherwise.

  \item \texttt{notice} (Boolean):  True if the file includes one and only one license notice, and possibly something else, not related to licensing  (in which case, \texttt{selse} will also be true). False otherwise.

  \item \texttt{multi} (Boolean): True if the file includes a multilicense text. False if \texttt{licen} is True, or \texttt{notice} is True, or the file includes no licensing information at all.

  \item \texttt{selse} (Boolean): True if the file includes something else besides licensing information, False if it includes only licensing information (license texts or license notices).

  \item \texttt{copy} (Boolean): True if the file includes at least one copyright notice. False otherwise.

  \item \texttt{swrong} (Boolean): Only when \texttt{licen} is True, True if ScanCode information is incorrect (w.r.t. manual file inspection by the authors). In this case, either the identified license(s) are not correct, or there are undetected licenses, or there are detected licenses that cannot be found in the file.

  \item \texttt{debian} (Boolean): \revchange{True if the document is a } file in the Debian copyright file format (in any version). We have found that a relatively large fraction of text files are in this format, and we consider it useful to label those as such.

  \item \revchange{\texttt{found} (String, in the \texttt{scancode} field format): Licenses that we have found in the file, but ScanCode did not. Only specified for plain text files with \texttt{licen} equal to True.}

  \item \revchange{\texttt{notfound} (String, in the \texttt{scancode} field format): Licenses that ScanCode identified in the file, but we have not found. Only specified for plain text files with \texttt{licen} equal to True.}
    
  \end{itemize}

\item \texttt{\bfseries truth\_uhash} (CSV file): \texttt{truth\_uhash.csv} is a file with the same columns as \texttt{truth}, but with only a representative for each family of documents with the same uhash. Uhashes are computed for text files by converting each document to lowercase, removing blanks, and removing copyright notices, and then computing SHA1 for the resulting string. This means that two text files with the same uhash have the same text, except for uppercase/lowercase, blanks and copyright notices. We have found that this selection of representatives for each uhash family is convenient for several analyses of the documents in the dataset.
\end{itemize}

 \section{Methodology and reproducibility}
\label{sec:methodology}

To produce the dataset, the following steps were taken:
\begin{enumerate}
\item \textbf{Selection and retrieval of documents} from \SWH. An intentionally broad selection criteria was designed and then executed to obtain a list of all documents that likely include the text of licenses or license notices. The result of this action is the Document Collection.
\item \textbf{Automated annotation} of the entire Document Collection. In order to make the dataset more useful, some metrics and characterizations were produced for each document in the collection. All of them are included in the dataset as CSV and JSON files (the Metadata Files).
\item \textbf{Manual annotation} of a large random sample. A random sample of the documents in the collection was manually annotated, producing the Annotated Sample.
\end{enumerate}

\begin{figure}
  \includegraphics[width=\textwidth]{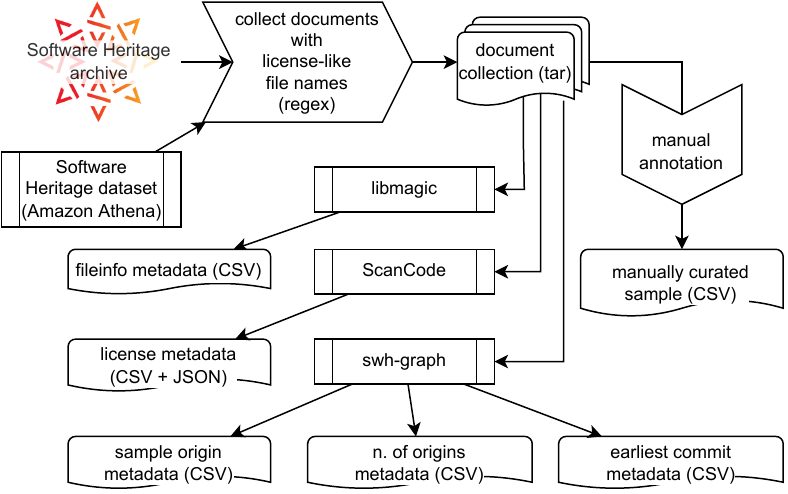}
  \caption{Dataset construction pipeline.}
  \label{fig:methodology}
\end{figure}

\noindent
Figure~\ref{fig:methodology} summarizes all these actions. The next subsections describe them in detail.

\subsection{Selection and retrieval of documents}

The first action was to select and retrieve all unique documents (file blobs) from \SWH that are likely to contain license texts or license notices. Unfortunately, there is no way of knowing for sure \emph{a priori} if a file contains a license, short of reading the file and finding one in it. Since this was not practical for the complete collection of documents in \SWH, we used an heuristic based on the filename.

This choice is based on a combination of convenience and strategy. Convenience, because SQL filtering based on filename is simple to perform using the preexisting \SWH graph dataset~\cite{swh-msr2019-dataset}. Strategy, because we expected it to capture most, almost all, licensing files: \revchange{it is a common development practice to advertise the licensing terms of a given software module using a file that adheres to well-established naming convention}. For example, in GitHub and GitLab it is recommended (and encouraged) to use files with names \texttt{LICENSE.txt}, \texttt{LICENSE.md} or \texttt{LICENSE.rst} so that they are easy to spot by humans and also detected automatically by tools. The OSI and the FSF have recommended to use filenames such as \texttt{LICENSE} or \texttt{COPYING} for the same purpose for decades.

Based on our experience with license files, we decided to use the following broad \revchange{file name} regular expression:
\begin{verbatim}
     ^([a-z0-9._-]+\.)?(copying|licen(c|s)(e|ing)|notice
     |copyright|disclaimer|authors)(\.[a-z0-9\._-]+)?$
\end{verbatim}
Using this expression, an SQL query was written\footnote{The complete SQL query is available as part of the dataset replication package~\cite{jesus_m_gonzalez_barahona_2023_8200352}, in the \texttt{replication-package.tar.gz} file.} to retrieve the SWHID, SHA1 checksum, and filename of all file blobs associated to at least one filename matching the above regular expression. Since a document may have been found by \SWH in many different repositories, with many different filenames, filenames retrieved with this query are very varied, although no document was retrieved if none of the filenames associated with it matched the expression. The SQL query was performed on the \SWH graph dataset~\cite{swh-msr2019-dataset} hosted on Amazon Athena (version \revchange{\DataEdIIILabel}).

The chosen regular expression is relatively lax and therefore matches many files containing data other than license texts or license notices. This was done on purpose, because while it is trivial to filter dataset blobs based on filenames using the \texttt{fileinfo} metadata (see Section~\ref{sec:datamodel}), it is cumbersome to \emph{extend} the dataset downstream to add all blobs of interest.

We then retrieved all selected blobs from the \SWH archive~\cite{swhipres2017} and archived them in a single \texttt{tar} file. This step was conducted in collaboration with the Software Heritage team, but can be independently replicated using any archive copy or mirror. This \texttt{tar} file is a materialization of the Document Collection, which is the starting point of our dataset.

\subsection{Automated annotation}

The whole collection of documents was later mined to gather various types of metadata, which we used to automatically annotate the whole dataset (see Figure~\ref{fig:metadata-schema}, discussed in Section~\ref{sec:datamodel}). The code we used for mining is available as part of the dataset replication package~\cite{jesus_m_gonzalez_barahona_2023_8200352}. The results of these actions are the Metadata Files.

To detect file MIME types and character encodings we used
\texttt{libmagic}~\cite{file-opengroup} on each document via the \texttt{python-magic}
Python bindings. For files with MIME type starting with \texttt{text/} and
UTF-8 encoding (or \emph{textual files} in the following for brevity) we also
computed line and word counts using custom Python code; for all files we
computed file sizes in bytes.

The licenses likely contained in each document have been detected by running the
ScanCode toolkit~\cite{scancode-toolkit} using its Python API. We run ScanCode
with no minimum score threshold---meaning that all detected licenses will be
returned, no matter the tool confidence in the result---and with a timeout of 2
minutes (per document).
ScanCode can perform many different types of analysis on (allegedly) license files.
We used both its license detection engine (corresponding to the \texttt{--license} command-line option) and its copyright notice detection engine (\texttt{--copyright}).
The most relevant results returned by ScanCode are included in the CSV file \texttt{blobs-scancode.csv.zst}, specifically: detected license and confidence score.
This allows to access detected license information easily by simply importing that file into any tabular analysis tool.
All additional information detected by ScanCode is available in the separate and much more detailed JSON file \texttt{blobs-scancode.ndjson.zst}.
Examples of information available only in the JSON file are: where within a file a given license detection rule matched; whether the detected license was a full text document or a notice, among others.
See the ScanCode documentation\footnote{\url{https://scancode-toolkit.readthedocs.io/}, accessed 2022-11-09} for complete details about the additional available information.

Finally, we used the compressed in-memory graph representation~\cite{saner-2020-swh-graph} of the Software Heritage archive to gather the \texttt{sample\_origins}, \texttt{nb\_origins}  and \texttt{earliest\_commit} metadata.
For \texttt{sample\_origins} and \texttt{nb\_origins} we used the \texttt{/leaves} API endpoint\footnote{\url{https://docs.softwareheritage.org/devel/swh-graph/api.html\#leaves}, accessed 2022-11-09} to traverse the transposed Merkle DAG of the archive and navigate from each document to all the origins referencing it.
For \texttt{sample\_origins} we just retrieved the first origin in the list (which is returned in an arbitrary order, so the selected sample origin is arbitrary as well); for \texttt{nb\_origins} we counted the number of origins.
\DataBlobNoOriginCount{} license documents ($\approx\,$\DataBlobNoOriginPct\% of the total) could not be mapped to an origin this way and lack origin metadata in the dataset.

For earliest commit information we used ad-hoc Java code (available as part of the replication package~\cite{jesus_m_gonzalez_barahona_2023_8200352}) to navigate the transposed graph from each document to all commits referencing it, which were counted as the number of occurrences of the document in the archive.
Then we selected the commit with the oldest timestamp among them and extracted its identifier and Unix time.

\subsection{Manual annotation}

For producing the Annotated Sample, we manually annotated a random sample of the Document Collection. We started by \revchange{sampling} \DocumentsManualCount{} random files from the whole collection of documents, which we will from now on refer to as the \emph{random subset}. We did that using specific seeds for a pseudo-random algorithm, so that the process can be easily replicated exactly if needed. We intend these files to be suitable for building corpus, training sets or a ground truth related to licensing information in files. 

We also computed normalized text for all plain text documents in the sample, by removing blanks, lowercasing, and removing one-line copyright notices, the SHA1 for the resulting text, to ease detection of documents with only ``cosmetic'' differences. With this, we found that we had \DocumentsManualNormalizedCount{} documents with different SHA1 for the normalized text (of a total of 6783 plain text documents).
\revchange{For comparison, for a sample size as large as the number of files in the Document Collection, a representative randomized sample with 99.9\% confidence level and 2\% margin of error would be of 6759 items (assuming p=0.5). Therefore, our sample, which is larger, has a confidence level higher than 99.9\% for a 2\% margin of error. In Subsection~\ref{subsec:findings-rq1} we will offer a more nuanced analysis of the confidence level and margin of error for the specific case of files with license text.}

We checked all documents in the random subset, whichever their format. We also did our best to identify texts in languages other than English, trying to find out license texts, license notices or copyright notices, by using automatic translators. For each of the documents, we annotated the data described in Section~\ref{subsec:description-sample}. It should be noted that all fields are Boolean, and that of the three first fields (\texttt{licen}, \texttt{notice} and \texttt{multi}) only one may be True.

The \texttt{licen} field is the most interesting one to identify full license texts, since it should include only one of them (plus maybe some other information not related to licensing). Similarly, \texttt{notice} is interesting to identify license notices. Files with one of \texttt{licen}, \texttt{notice}, or \texttt{multi} set to True could be used if there is interest in files with any kind of licensing information. These fields can be used in combination with the \texttt{selse} field when it is important to exclude files with other information unrelated to licensing.

The \texttt{copy} field is intended to help in checking heuristics for detecting copyright notices. The \texttt{debian} field intends to ease the detection of likely Debian copyright files, which could be of interest given its popularity in the dataset, and its formatting rules which in most cases allow for automatic parsing.

To clarify meanings, we consider:
\begin{itemize}
  \item A ``license text'' to be the complete text of a license, maybe with a disclaimer, and maybe with one or more copyright notices. This would be, for example, the text of the MIT, Apache 2.0 or GPL-3.0 licenses, but also a text such as ``This software is put in the public domain and you can do whatever you may want with it'' or ``All rights reserved, to use this software contact company XXX''. 
  \item A ``license notice'' to be a notice stating which one is the applicable license, but not including the text of the license itself. For example, texts such as ``This software is distributed under the terms of the Apache 2.0 License'', maybe accompanied by links to the complete license, copyright notices and disclaimers. 
  \item A ``copyright notice'' to be a notice stating who is the copyright holder. An example would be ``Copyright 2020 Free Software Foundation'', but also many other variations of this text, with our without the copyright symbol.
\end{itemize}

\begin{table}
  \center
  \caption{Cohen's Kappa values, before and after the discussion between the two annotators, for main annotation fields in the Annotated Sample. Those values were calculated on a subset of 399 files. Cohen's Kappa is calculated for binary values for \texttt{licen}, \texttt{notice}, \texttt{multi}, \texttt{swrong}. ``ftype'' is used to compute Cohen's Kappa for the type of file (license text, license notice, multilicense, or none of them) as a categorical value.}
  \begin{tabular}{l|r|r}
    \textbf{Field}  & \textbf{Before discussion} & \textbf{After discussion}\\ \hline
    licen &  0.762 & 0.944 \\
    notice & 0.714 & 0.803 \\
    multi &  0.672 & 0.818 \\
    swrong & 0.496 & 0.590 \\
    ftype &  0.756 & 0.898 \\
  \end{tabular}
  \label{tab:annotated-cohen}
\end{table}

\noindent
\revchange{
  For determining the value of all these fields, two authors went manually through all licenses in the dataset, filling in values. After that, they discussed those cases where they were not in agreement, and fixed those cases where any of them detected an error, and thus they were in agreement. The remaining cases were decided by reaching consensus with a third author, to produce a single final version. The detailed criteria, and other details about the definitions used, can be found in the manual validation notebook available in the replication package~\cite{jesus_m_gonzalez_barahona_2023_8200352}. We have estimated the effort needed for the initial annotation to be about 1 hour for each batch of 100 files (with different \texttt{uhash}) for each of the annotators; hence about 80 hours/annotator in total. The notebooks used for assisting in the annotation can also be found in the replication package.
}

\revchange{
  The values for agreement between the first two annotators are shown in Table~\ref{tab:annotated-cohen}. It is important to notice how the discussion phase significantly increased agreement between annotators. This was both because it helped to find errors in the annotation, but also because the conversation helped to agree on how to deal with corner cases. It is also worth noticing the different levels of agreement reached for the different fields. For deciding which type of file (license text, license notice or multilicense), the level of agreement reached before the consensus phase is high. However, the disagreement is higher when deciding if ScanCode was wrong or not. This is due in part to the fact that criteria for deciding on file type were more precise, and in part because the labels produced by ScanCode are not very convenient for our study. ScanCode is targeted to find any hint of a license (be it the whole license, a license notice, or even just a reference to the license name). Because of that, some of its labels could be considered false positives or not depending on how strict we are. In addition, some of the labels are a bit ambiguous, such as ``LicenseRef-scancode-warranty-disclaimer'', which could be considered as a hint for a new license or just as a part of a license.
}

 \section{Characterization of the dataset}
\label{sec:characterization}

This section presents what can be found in the dataset, by characterizing it from several points of view.

\revchange{Note that all SWHID references in footnotes in the following are clickable hyperlinks in the electronic version of this paper. Independently from that, they can all be resolved by visiting {\small \url{https://archive.softwareheritage.org/<SWHID>}} to inspect the mentioned license document.}

\subsection{Main parameters of documents}
\label{sec:statistics}

\begin{figure}
  \centering
  \includegraphics[width=\textwidth]{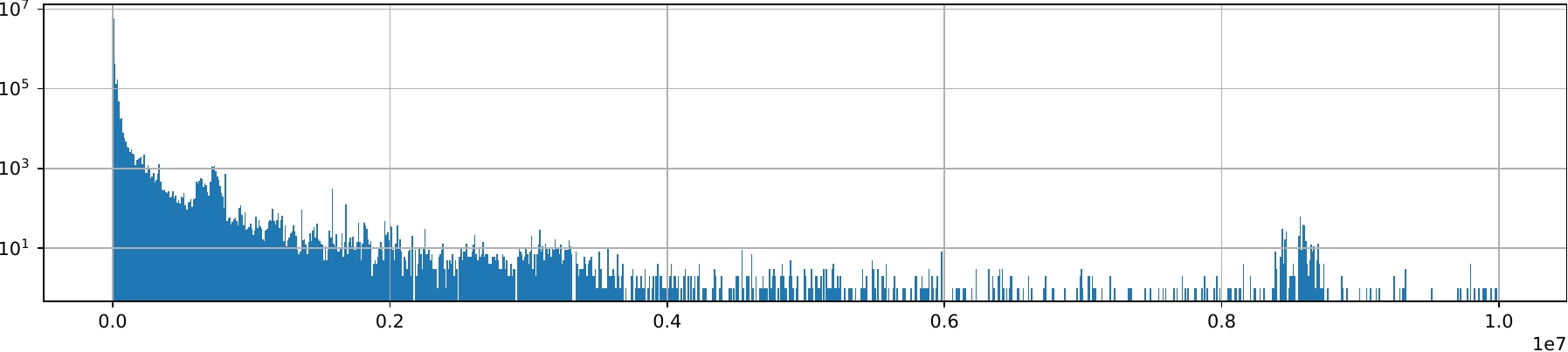}
  \caption{Histogram of all documents smaller than 10 MiB, logarithmic scale, 1000 bins.}
  \label{fig:documents-size}
\end{figure}

The documents in the dataset are of many different sizes. The largest one\footnote{SWHID \swhid{swh:1:cnt:36406a1eee032e80a284d3ed9f5176bba67be064}} is a binary file of about \num{104} MiB; the largest text file\footnote{SWHID \swhid{swh:1:cnt:cdc98c898b1d257ddb4752ee7a1c85ed3ddf5673}} is of about \num{82} MiB, and is a very large JSON file. But most files are much smaller. Figure~\ref{fig:documents-size} shows a histogram for all documents smaller than 10 MiB (note the logarithmic scale). As can be seen, even when there are documents of all sizes, the smaller the file, the more documents of that size.

This is confirmed in Table~\ref{tab:size-distribution}, which shows the description of the collection in terms of size, both for all documents, and only for those of type \texttt{text/plain}.

\begin{table}
  \centering
  \caption{Descriptive statistics of the size of documents in the collection.} 
  \label{tab:size-distribution}
  \begin{tabular}{rrrc}
                    & \textbf{All documents} & \textbf{\texttt{text/plain} documents} & \\ \hline \hline
\textbf{Count}      & 6.859 million & 5.721 million & documents \\ \hline
\textbf{Mean}       & \num{10159}   & \num{6658}    & bytes \\ \hline
\textbf{Std~dev}    & \num{245021}  & \num{133646}  & bytes \\ \hline
\textbf{25\%}       & \num{1065}    & \num{1064}    & bytes \\ \hline
\textbf{50\%}       & \num{1080}    & \num{1075}    & bytes \\ \hline
\textbf{75\%}       & \num{2241}    & \num{1320}    & bytes \\
  \end{tabular}
\end{table}

\begin{figure}
  \includegraphics[width=\textwidth]{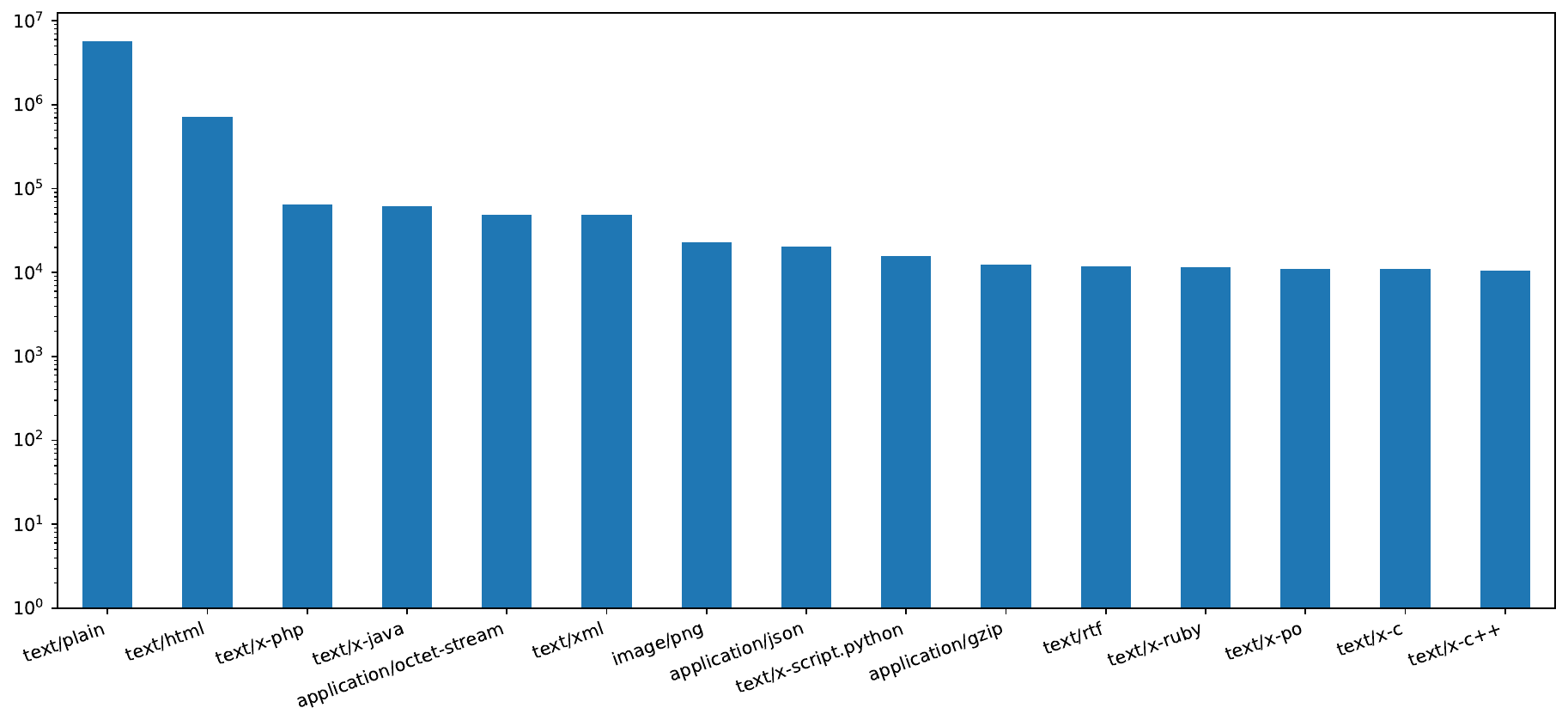}
  \caption{Top filetypes in the collection, by number of documents, logarithmic scale.}
  \label{fig:documents-filetypes}
\end{figure}

The count of documents by filetype is also interesting. Figure~\ref{fig:documents-filetypes} offers a bar chart for the main filetypes (note the logarithmic scale). It shows how the most represented type, by almost two orders of magnitude, is \texttt{text/plain} (plain text). \DataMimeTextPlainPct\% of the corpus blobs are \texttt{text/plain} and \DataMimeTextPct\% \texttt{text/} of some kind (including HTML, XML, and LaTeX). Other interesting (small) classes are rich text formats like RTF, image files, and PDFs. We have manually verified that at least some of these are actually used to distribute licensing terms, the rest is a small amount of noisy data. \revchange{Some examples of these marginal classes, just as a curiosity: a RTF file,\footnote{SWHID \swhid{swh:1:cnt:2e26bf237427aaa56f99846acb1aeb94198119e9}} an image (not exactly a license, but a Creative Commons logo that could be interpreted as a license),\footnote{SWHID \swhid{swh:1:cnt:606a3bce98a4ade7d80c2761b8458d79438a3c6f}} and a PDF file.\footnote{SWHID \swhid{swh:1:cnt:78ec4db8002adeae4fcbfa5f56b3c1e51bfaf8c5}}}

Word frequencies in the collection may also help to better understand the dataset. Table~\ref{tab:top-words} shows the most used words in the documents of the collection, after removal of English stopwords and single-character tokens (the exact procedure is presented as an example of dataset usage, including code, in Section~\ref{subsec:usage-collection}). Most of these words correspond to meaningful terms in the semantic domain of open source licensing, which is an indication that the dataset actually includes many documents related to software licensing. \revchange{Just as a curiosity, the term ``nasl'' which appears in the list, is found in files which are inventories of scripts in the NASL scripting language,\footnote{\url{https://en.wikipedia.org/wiki/Nessus_Attack_Scripting_Language}} commonly used in network appliances. In them, ``.nasl'' appears as a file name extension which, due to word tokenization, are identified as words. For example, one of these inventories includes the name of more than \num{65 000} of these files.\footnote{SWHID: \swhid{swh:1:cnt:c7f43dd49cbedb819fc247b3bfe5ae45841738dc}} We can consider it as a false positive, since they are not really words in licenses, but we kept it in the list for the sake of transparency on the characteristics of the dataset.}

\begin{table}
  \caption{Top-15 words in the license corpus by frequency.}
  \label{tab:top-words}
  \centering
\begin{tabular}{c|r}
    \textbf{Word} & \multicolumn{1}{c}{\textbf{Frequency}} \\
    \hline
    \begin{filecontents*}{words_top.csv}
word,frequency
software,73848600
license,65120569
copyright,53813792
use,36492256
work,35606158
without,27946481
including,27065320
source,25432478
notice,25007288
conditions,24425279
shall,22558504
provided,22523413
gplv2,21709671
nasl,20805400
following,19333506
copy,18296628
may,18092010
com,17343283
permission,16720683
must,16454459
code,15776625
rights,15413909
warranties,15202030
implied,15025359
liability,14817590
terms,14639581
form,14191924
limited,14157137
damages,13337383
purpose,13227506
\end{filecontents*}
\csvreader[
      head to column names,
      range=-15,
    ]{words_top.csv}{}{
      \word & \num{\frequency} \\
    }
  \end{tabular}
\end{table}

\subsection{Licenses found}
\label{subsec:characterization-licenses-found}

To facilitate a preliminary quantification of the number of documents with licensing information we used the results of running the ScanCode tool on all \texttt{text/plain} documents of the collection, available in the \texttt{scancode} CSV file in the Metadata Files. The first result of this analysis is that ScanCode identified some licensing information in \num{4859282} documents (70.84\% of the total number of documents in the collection). 

\begin{figure}
  \center
  \includegraphics[width=0.75\columnwidth,trim=1.5cm 1.5cm 1.5cm 1.5cm,clip]{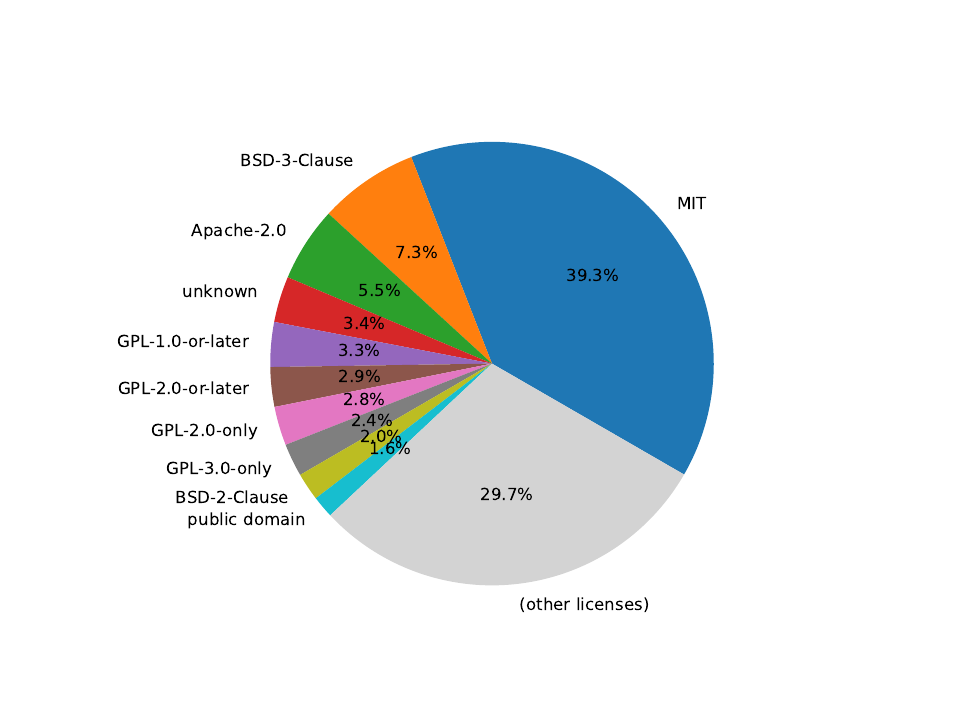}
  \caption{Top licenses in the Document Collection, by number of documents in which they were detected by ScanCode, calculated over the total number of license detections in all documents.}
  \label{fig:top-licenses}
\end{figure}

The \texttt{scancode} CSV file lists license texts or notices that ScanCode found in each document. Therefore, we can find out in how many documents each license was found. Using this data, we have produced Figure~\ref{fig:top-licenses}, which shows the list of the main licenses by number of documents in which they were found. It shows that MIT is the most prevalent open
source license variant in the corpus, followed by 3-clause BSD, and Apache-2.0. Considering that we are counting \emph{license variants} here, having MIT and BSD at the top makes intuitive sense, because their text usually includes a copyright notice which needs to be instantiated by individual authors for each different project.

\revchange{In other words this should not be interpreted as a measure of license \emph{popularity}, but of license \emph{variability}}.

\subsection{Occurrence of known licenses}
\label{subsec:characterization-known-licenses}

To learn about how comprehensive the Document Collection is, we checked if already known licenses are present in the collection. To make the check a reliable lower bound, we only checked for licenses which are exactly as in a well known list of software licenses. As a well known list, we used the ScanCode LicenseDB collection, which to our knowledge is the largest publicly available collection of licenses. It includes all of OSI-recognized licenses, but also many others. The total number of license texts in it is 1879.

For checking the occurrence of these license texts in the Document Collection, we computed the SHA1 checksum for each of them, and searched for it in the Metadata Files. Using \texttt{license\_blobs} we obtain the SWHID for each SHA1, and then, searching it in \texttt{nb\_origins}, we get the number of repositories (origins, in \SWH parlance) in which the document was found. For a total of 1879 license texts in the ScanCode LicenseDB, we found 1847 documents with exactly the same text (98.29\%).

Table~\ref{tab:licenses-scancode-occurrences} shows the licenses with more occurrences in different repositories which are in the ScanCode LicenseDB and can also be found in the Document Collection with exactly the same text. This list shows some of the more popular FOSS licenses, such as versions of the GPL or Apache licenses. But also some much less well known licenses, such as the Ubuntu Font License, the Boost License (for a C++ library) or Unlicense (a minimalist license similar to MIT). It is important to note that this list does not imply by any mean that these licenses are more popular than others, only that they are present with exactly the same text in that number of repositories. For example, the MIT or the BSD licenses are low in this list because they are usually included with modifications, such as a specific copyright line which is different from repository to repository.

\begin{table}
  \caption{Top licenses in ScanCode LicenseDB that can be found in the Document Collection, by number of occurrences in different repositories. License names are those of the corresponding file in ScanCode LicenseDB.}
  \label{tab:licenses-scancode-occurrences}
  \centering
  \begin{tabular}{l|r}
    \textbf{License} & \textbf{Repositories} \\ \hline
    gpl-2.0 & \num{1216365} \\
    gpl-3.0 & \num{903164} \\
    cc0-1.0 & \num{237457} \\
    apache-2.0 & \num{165453} \\
    ubuntu-font-1.0 & \num{159923} \\
    boost-1.0 & \num{155654} \\
    unlicense & \num{109687} \\
    lgpl-3.0 & \num{94745} \\
  \end{tabular}
\end{table}

\revchange{
We also checked what licenses appear \emph{together} in the same file more frequently. ScanCode identifies as many licenses in a file as it can, producing a list of license identifiers for each of them. We analyzed this list for all files with more than one license identified by ScanCode. The top license pairs found are shown in Table~\ref{tab:licenses-scancode-together}.
}

\begin{table}
  \caption{Top license pairs, by number of files in which they appear together in the same file, as identified by ScanCode (excluding licenses that ScanCode could not identify).}
  \label{tab:licenses-scancode-together}
  \centering
  \begin{tabular}{l|r}
    \textbf{License pairs} & \textbf{Files} \\ \hline
    Apache-2.0 \& MIT & \num{99774} \\
    GPL-1.0 \& GPL-2.0 & \num{74706} \\
    GPL-2.0 \& MIT & \num{59880} \\
    BSD-3-Clause \& BSD-2-Clause & \num{53764} \\
    BSD-3-Clause \& MIT & \num{53034} \\
    Apache-2.0 \& BSD-3-Clause & \num{52662} \\
    GPL-1.0 \& GPL-3.0 & \num{52030} \\
  \end{tabular}
\end{table}

\subsection{Description of the Annotated Sample}
\label{subsec:characterization-annotated-sample}

\begin{figure}
\includegraphics[trim=3cm 1cm 3cm 1cm,clip,width=0.5\columnwidth]{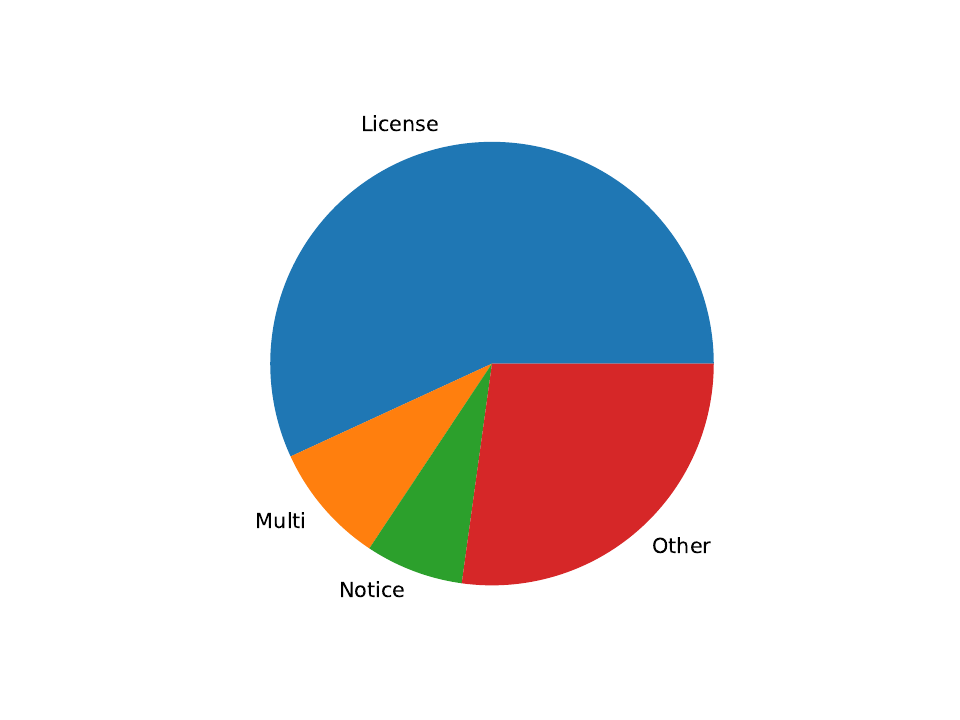}
  \raisebox{3cm}{\begin{tabular}{l|r|r}
    \textbf{Kind} & \textbf{Documents} & \textbf{Fraction} \\ \hline \hline
    License & 4608 & 56.87\% \\
    Multilicense & 713 & 8.80\% \\
    Notice & 579 & 7.14\% \\
    Other & 2202 & 27.17\% \\ \hline
    Total & \DocumentsManualCount & 100.00\% \\ 
  \end{tabular}}
  \caption{Annotated Sample: kinds of documents. ``License'' are documents including the complete text of a single license. ``Notice'' are documents including a single license notice. ``Multilicense'' are documents including the whole text of licenses, and/or license notices (more than one). The three categories are disjoint. ``Other'' are documents that do not include complete licenses or license notices. The total fraction of documents with licensing information (``Licenses'', ``Multilicense'' and ``Notice'') is 72.81\%.}
  \label{fig:truth-kinds}
\end{figure}

The Annotated Sample is a subset of \DocumentsManualCount{} documents, selected randomly from the  Document Collection. The main value of the sample is the manual characterization of documents according to the licensing data they have, if any. As we detailed in Section~\ref{subsec:description-sample}, we classified every document according to the kind of licensing information we found in it: full text of a license, a license notice, more than one license text or license notice, or none of them. Figure~\ref{fig:truth-kinds} shows how many documents of each kind were found in the sample. While manually annotating licenses, we also identified which ones included copyright notices, or were in the Debian copyright file format. Table~\ref{tab:truth-characteristics} offers details about these additional characteristics.

It is also worth noticing that we have found, during this manual annotation, that 121 documents containing the full text of a license were wrongly identified by ScanCode: either no license was found, despite the document having the full text of one, or the license identified does not correspond with the license in the document. This amounts to 2.7\% of the total number of documents with full-text licenses in the sample, in plan text format. Therefore, we estimated we could use ScanCode for characterizing licenses in the sample, as we do later in this section, and for the whole collection, as we do in Section~\ref{sec:findings}.

\begin{table}
  \caption{Annotated Sample: Documents with characteristics of interest. ``Debian format'' means the document is in the Debian copyright file format. ``Copyright notices'' means the document includes at least one copyright notice.}
  \label{tab:truth-characteristics}
  \centering
  \begin{tabular}{l|r}
      \textbf{Characteristic} & \textbf{Documents} \\ \hline
      Debian format & 294 \\
      Copyright notices & 4608 \\
  \end{tabular}
\end{table}

Being a random sample of the whole collection, the statistics of the sample are relatively similar to the whole collection, although obviously with less variation (the sample includes approximately one in every 1000 documents in the collection). Some descriptive statistics of the Annotated Sample are shown in Table~\ref{tab:sample-size-distribution}.

\begin{table}
  \caption{Descriptive statistics of the size of documents in the annotated sample.} 
  \label{tab:sample-size-distribution}
  \centering
  \begin{tabular}{rrrc}
            & \textbf{All documents} & \textbf{\texttt{text/plain} documents} & \\ \hline \hline
\textbf{Count}      & \num{\DocumentsManualCount}   & \num{6783}   & documents \\ \hline
\textbf{Mean}       & \num{12 910}  & \num{8 382}   & bytes \\ \hline
\textbf{Std.~dev.}  & \num{27 0847} & \num{207 377} & bytes \\ \hline
\textbf{25\%}       & \num{1 065}   & \num{1064}   & bytes \\ \hline
\textbf{50\%}       & \num{1 079}   & \num{1075}   & bytes \\ \hline
\textbf{75\%}       & \num{2 265}   & \num{1316}   & bytes \\
\end{tabular}
\end{table}

\begin{figure}
  \center
  \includegraphics[width=\columnwidth,trim=3.5cm 0 3.5cm 0]{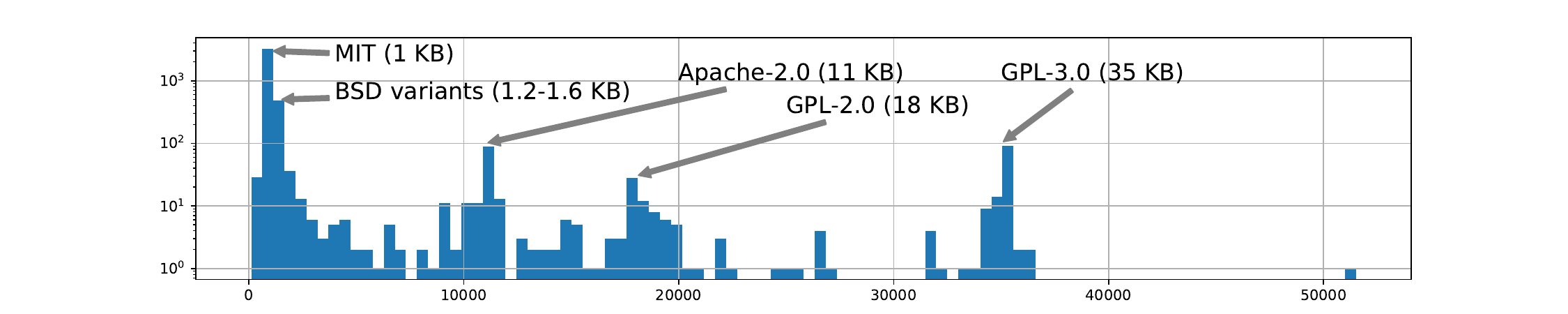}
  \caption{Annotated Sample: Histogram of documents identified as containing the text for a single license, by size (logarithmic scale, size in bytes). Labels show the approximate length of some licenses, pointing to the corresponding peaks in the histogram.}
  \label{fig:truth-licen-text-size}
\end{figure}

Focusing on documents which include the text of a single license, Figure~\ref{fig:truth-licen-text-size} shows a histogram of their size. As it seems reasonable, these files are much shorter than those containing license texts, with those that are longer being less than 1700 bytes, and the large majority of them being below 1\,KiB. We also include next to the figure a table with the most frequent licenses, with their approximate length (which varies depending on how exactly it was formatted and complemented with other information in the document). It is interesting to see how this size matches some of the peaks of the histogram. \revchange{For comparison, Figure~\ref{fig:truth-licen-text-size} shows the number of files in buckets for different file sizes, also accompanied with the sizes of some common licenses.}

To complement this data, Table~\ref{tab:truth-licen-scancode} shows the list of documents in the Annotated Sample by license, among the documents containing the text of a single license. The identification of licenses, in this case, was done by ScanCode. This table shows immediately how variants of the MIT license are a very large fraction of all documents. Variants of BSD and Apache 2.0 licenses are also prominent. This means that software authors tend to write these licenses with more variations than other, also popular licenses. 

\begin{table}
  \center
  \caption{Annotated sample: top licenses, as identified by ScanCode, among files annotated as containing the text for a single license.}
  \begin{tabular}{l|r}
    \textbf{License} & \textbf{Documents} \\ \hline
    MIT          & 3282 \\
BSD-3-Clause     & 230 \\
Apache-2.0       & 123 \\
GPL-3.0-only     & 107 \\
BSD-2-Clause     &  88 \\
GPL-2.0-only     &  38 \\
ISC              &  36 \\
Others           & 296 \\
  \end{tabular}
  \label{tab:truth-licen-scancode}
\end{table}

\begin{figure}
  \center
  \includegraphics[width=\columnwidth,trim=3.5cm 0 3.5cm 0]{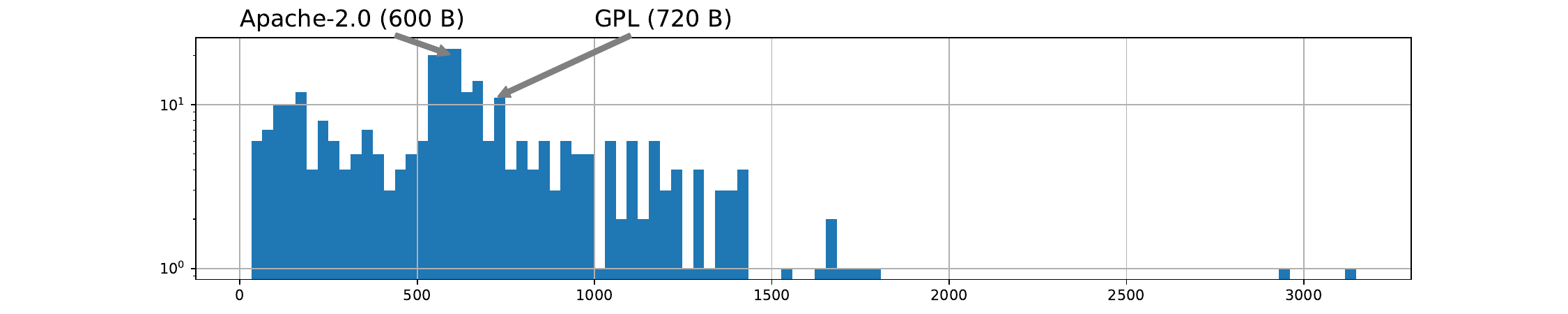}
  \caption{Annotated Sample: Histogram of documents identified as containing a single license notice, by size (logarithmic scale, size in bytes). Labels show the approximate length of some common license notices, pointing to the corresponding peaks in the histogram.}
  \label{fig:truth-licen-text-size-single}
\end{figure}

Table~\ref{tab:truth-notice-scancode} shows the list of documents by license, among the documents annotated as containing a single license notice. The license identification was done by ScanCode. In this case, there is little representation of the ``short'' licenses (such as MIT or BSD), since they are usually included verbatim, and only in rare circumstances as a notice. Larger licenses, such as versions of the GPL or Apache, are more popular in this set, since it is common practice to include a license notice pointing to the complete text somewhere else.

\begin{table}
  \center
  \caption{Annotated sample: top licenses, as identified by ScanCode, among files annotated as containing a single license notice.}
  \begin{tabular}{l|r}
    \textbf{License}      & \textbf{Documents} \\ \hline
    Apache-2.0       & 73 \\
GPL-3.0-or-later     & 26 \\
GPL-2.0-or-later     & 25 \\
GPL-2.0-only         & 12 \\
MIT                  & 11 \\
Others               & 147 \\
  \end{tabular}
  \label{tab:truth-notice-scancode}
\end{table}

\revchange{
  The kind of analysis and annotation we have done is not designed for evaluating the correctness of ScanCode, because we focused mainly on classifying files for facilitating further processing. However, we already showed how the error rate of the tool is rather small. In Table~\ref{tab:truth-scancode-found} and Table~\ref{tab:truth-scancode-notfound} we show the main reasons for those errors: (i) the licenses we found in the files, but ScanCode could not find, and vice versa (ii) the licenses ScanCode ``detected'', but we could not find in the corresponding files.
}

\begin{table}
  \center
  \caption{Annotated sample: top licenses found in our manual analysis that were not found by ScanCode, for plain text files identified as containing only a single license text. ``No ScanCode Id'' means we could not find a ScanCode Id for it, ``License Not Identified'' means that ScanCode found a license, but could not recognize it.}
  \begin{tabular}{l|r}
    \textbf{License}      & \textbf{Documents} \\ \hline
    No ScanCode Id       & 42 \\
    License Not Identified & 3 \\
    MIT & 2 \\
    BSD-3-clause         & 2 \\
    Others               & 14 \\
  \end{tabular}
  \label{tab:truth-scancode-found}
\end{table}

\begin{table}
  \center
  \caption{Annotated sample: top licenses that were identified by ScanCode, but we couldn't find in the file, for plain text files identified as containing only a single license text. ``License Not Identified'' means that ScanCode found a license, but could not recognize it.}
  \begin{tabular}{l|r}
    \textbf{License}      & \textbf{Documents} \\ \hline
    Unicode       & 10 \\
    License Not Identified & 3 \\
    Proprietary License & 2 \\
    GPL 1.0 or later    & 2 \\
    Others              & 61 \\
  \end{tabular}
  \label{tab:truth-scancode-notfound}
\end{table}

 \section{Findings}
\label{sec:findings}

In this section we present the answers to the two research questions posed in the Introduction, and also a list of examples of curious and interesting license texts that we found during our manual analysis.

\subsection{RQ1: How many distinct licenses does the dataset contain?}
\label{subsec:findings-rq1}

\revchange{For this RQ we are interested in learning how many of the files in the Document Collection contain the full text of a single license. Since each file in the Collection is (by construction of the dataset) different, the number of these files will be the number of distinct license texts (with very small or very large variations between them).}

\revchange{
  We answer this question by manually annotating the random subset. We read all documents in this sample, and assessed which ones corresponded to the full text of a license and contained no other licensing information. As presented in Section~\ref{subsec:characterization-annotated-sample}, the fraction of files with the full text of a single license is 56.87\%. If we consider this as a binomial distribution (with 1 for \texttt{licen}=True and 0 for \texttt{licen}=False), we can compute its mean (0.5687) and its standard deviation (0.4952). With these numbers, we can compute the margin of error for a confidence level of 99\% (using 2.575 as z-value). The resulting margin of error is 0.0107, or 1.07\%. In other words, we can estimate that the number of files composed by the text of a single license is 56.87\%, with a margin of error of 1.07\% (see Table~\ref{tab:documents-single-license}).
}

\begin{table}
  \caption{Documents with a single license text in the sample and in the whole collection, with estimation of error (99\% confidence level).}
  \label{tab:documents-single-license}
  \centering
  \begin{tabular}{l|r|r}
    \textbf{Set}   & \textbf{Total documents} & \textbf{Documents with single license text}  \\ \hline
    Sample & \DocumentsManualCount & $56.87\%\pm1.07\% (4608\pm87)$ \\
    Collection & 6.859 million &  $56.87\%\pm1.07\% (3.900 \mbox{ million} \pm \num{73391})$ \\
  \end{tabular}
\end{table}

It is important to notice that this result is a lower bound of the number of full-license documents ever published. We know for sure that the number could be larger, because we also have (in our collection) full-text licenses in multi-license documents. But we have not analyzed yet how many other different full-text licenses are in them. And, of course, there could be some other different full-text licenses in documents not in our collection. This can be either because they are in \SWH documents with filenames not captured by our query, or because they are not archived in \SWH.

Compared to the about 500 license texts recognized by SPDX, or the 2000 license texts in the ScanCode LicenseDB, the number of full-text license variants in our collection is huge. With RQ2 we explore some of the reasons leading to this diversity.

\subsection{RQ2: Why are there so many distinct licenses in the dataset?}

We answer this question using several approaches:

\begin{enumerate}
\item \emph{Manual inspection of the sample}. While inspecting the documents in the Annotated Sample, we found that many documents were similar to the ``canonical'' licenses (e.g., those recognized by SPDX). In particular, we found a very large quantity of variants of the MIT license, which in many cases only differ in copyright notices: due to the structure of the license, it is usual that a license file with a mention of the MIT license includes one or more copyright notices. This is the cause of many slightly different variants of the license text.

Slight changes of the license text happen for all licenses, but it happens less frequently in longer licenses, such as the GPL versions. One reason is that the GPL license is usually stored in a single file, with copyright notices in other files; this is also the practice recommended by the license itself. In shorter licenses this is not what we have found. For instance, the text of the MIT license (and also of other shorter licenses, such as BSD) is small enough for being included together with copyright notices.

  We quantified the number of variants using ScanCode, and found 3031 variants of the MIT license in our sample out of a total of 4248 licenses---or 71.35\% of the total variety of documents. The next license with more variants is BSD which, adding variants of BSD-2-clauses and BSD-3-clauses together, gives 288, or 6.77\%. In contrast, GPL-2 and GPL-3 together give only 128 documents, or 3.01\%.

  \revchange{Just as an illustration of the many ways in which the MIT (and other) license can be found, consider the content of three different files in our Annotated Sample: (i) the MIT license text, in a file with a heading in Japanese, and a list of credits;\footnote{SWHID \swhid{swh:1:cnt:9ea952f4a37478f17f2a2aafb45ced7a4df67de2}} (ii) an HTML template including the MIT license text;\footnote{SWHID \swhid{swh:1:cnt:aa3157cb23f7de5d062ab5d0bf0ffb44bb719df9}}; and (iii) the MIT license as quoted text.\footnote{SWHID \swhid{swh:1:cnt:509b6082ee6debe85c005d80f047668d70dd1cb8}}}
  
\item \emph{Normalization of the text in documents.} We decided to check if the hypothesis that copyright notices are the cause of a large number of documents by running documents through some normalization heuristics. These heuristics remove most copyright notices found in a single line (i.e., lines with variants of the word ``copyright'', maybe a list of words, and maybe a year, as in ``Copyright 2012 The Foo Project Developers.''). These heuristics also normalize to lowercase, and remove blank characters. Using it, we found that the total number of different documents in the annotated sample was reduced to \DocumentsManualNormalizedCount, from a total of \DocumentsManualCount{} (53.94\%).

  Extending these results to the whole collection, that would mean a reduction from about \DataBlobCountApproxM\, million documents to about 3.7 million documents. These heuristics, by construction, reduce diversity mostly on documents with license text. Taking this into account, together with our previous estimation of the total number of files with a single license text as 3.9 million, we can state that variations in upper/lowercase and blanks in copyright statements are the major cause of variety in documents with license texts.

\item \emph{Automatic annotation with ScanCode}. ScanCode checks for patterns in the text that identify licenses. The fact that ScanCode identified accurately a very large fraction of licenses means that those patterns worked well, and that those license texts identified should be similar to the ``canonical'' licenses. \revchange{Indeed, 70\% of the licenses in the dataset are identified by ScanCode with a score of 100 in a scale 0-100; only 15\% with a score lower or equal to 90; the average detection score across the dataset is 93.} Again, this goes in the direction of small variations (in copyright notices, blanks, etc.) being the main reason of license diversity.
\end{enumerate}

\noindent
Summarizing, we offer evidence that single-line copyright notices, blanks and upper/lowercase are a very important cause of the diversity of documents having full-text licenses. However, still more work is needed to find out exactly in which cases documents are really different licenses, not different only in ``cosmetic'' aspects, but in the actual wording of the license text.

\revchange{
It is also important to highlight that, according to Figure~\ref{fig:truth-kinds}, 27.70\% of the data in the Annotated Sample, and probably hence, in the entire dataset, do not contain licensing information. This is not only an additional reason for having so many different files in the dataset, but also something to take into account when using the dataset. For most analysis on it, all those ``other'' files should be excluded, or when that is not possible, kept in mind when discussing results.
}

\subsection{Curious and interesting licenses found}

While manually studying licenses for the Annotated Sample, we found many interesting cases. Below we present a non-exhaustive sample of these cases, which will help to better understand the kind of documents present in the collection.

\begin{figure}
  {\footnotesize
\begin{verbatim}
# -*- coding: utf-8 -*-
# vim: autoindent shiftwidth=4 expandtab textwidth=120 tabstop=4 softtabstop=4
 
###############################################################################
# OpenLP - Open Source Lyrics Projection                                      #
# --------------------------------------------------------------------------- #
# Copyright (c) 2008-2018 OpenLP Developers                                   #
# --------------------------------------------------------------------------- #
# This program is free software; you can redistribute it and/or modify it     #
# under the terms of the GNU General Public License as published by the Free  #
# Software Foundation; version 2 of the License.                              #
\end{verbatim}
}
\caption{First lines of a file with some editor commands.}
  \label{fig:cases-editor}
\end{figure}

\begin{figure}
  {\footnotesize
\begin{verbatim}
                    GNU GENERAL PUBLIC LICENSE
 
 MarginBot is Copyright (C) 2014 Howard Fenter III.
 If you want to help further development of this code,
 Please send a donations to: 
 Bitcoin: 1LtVC2TE88b9zJcf6NFk4fzupM74QGUXQB
 Litecoin: LgKWYe7uisDkfz2LDeYi7tKEHukJdoziyp
 For any questions please send e-mail directly to marginbot@fuckedgox.com
 
 Commercial Licensing for MarginBot
 is available by contacting me at: marginbot@fuckedgox.com
 
 You may use, distribute and copy MarginBot under the terms of
 GNU General Public License version 3, which is displayed below.
 
-------------------------------------------------------------------------
 
                   GNU LESSER GENERAL PUBLIC LICENSE
                       Version 3, 29 June 2007
\end{verbatim}
  }
  \caption{First lines of a file with LGPL-3.0 license, with some notes for donation and offer of a commercial license.}
  \label{fig:cases-donation}
\end{figure}

\begin{figure}
{\footnotesize
\begin{verbatim}
Copyright (C) 2009 James Pike. All rights reserved. This is subject to
change in the future.
\end{verbatim}
  }
  \caption{Example of a proprietary license.}
  \label{fig:cases-proprietary}
\end{figure}

\begin{figure}
{\footnotesize
\begin{verbatim}
fileFormatVersion: 2
guid: 5e1ce912f300f604aaa78c66058b22b8
timeCreated: 1452687307
licenseType: Free
DefaultImporter:
  userData: 
  assetBundleName: 
  assetBundleVariant:
\end{verbatim}
  }
  \caption{Example of a license type just stating ``free''.}
  \label{fig:cases-free}
\end{figure}
  
\begin{itemize}
\item Very long lists of attributions, with no license information, detected by ScanCode as ``public domain''.\footnote{SWHID \swhid{swh:1:cnt:f961852cee6ee9e9a0b8a25af5d090ddb6abe6a8}} This document shows how difficult it is for heuristics to detect the right license, or even that a document includes a license.
\item A license notice with editor settings (a so-called Vim ``modeline'') in the first line (see Figure~\ref{fig:cases-editor}).\footnote{SWHID \swhid{swh:1:cnt:711ded4ae27c43ba18a71ad05e9466a268e4387a}} This document shows how varied the spurious information that a document with licensing information may have is.
\item Truncated text of GPL-2.0.\footnote{SWHID \swhid{swh:1:cnt:46ae7b2bee342168dc48d6ca7fa1753b98e525d8}} This document includes only the beginning of the GPL-2.0 license, apparently because of some error. It shows how difficult it may be to single out licenses (most automated tools would signal this as a new license, based on GPL-2.0).
\item LGPL-3.0 license, with some notes for donation and offer of a commercial license (see Figure~\ref{fig:cases-donation}).\footnote{SWHID \swhid{swh:1:cnt:62319023a68b04f23ea30931bb1a7c1a3e741fba}} Those notes do not really change the license (which remains LGPL-3.0), but for an automated tool, it may be really difficult to identify this is not a variant of the LGPL. In fact, ScanCode identifies it as LGPL-3.0-only, but also as GPL-1.0-or-later and GPL-2.0-only.
\item Specific license, apparently written by authors of some software to suit their needs.\footnote{SWHID \swhid{swh:1:cnt:eb9ed7bfc458af9796b59426d54d0f97a199078f}} This seems to be the case of a legit license, different to the ``usual'' licenses, but still a license that should be identified as such.
\item Test for an end-user (proprietary) license agreement.\footnote{SWHID \swhid{swh:1:cnt:b864764d9fc4d55eb09e123e42ede11519556d18}} Apparently, does not allow redistribution or publication of the source code, which means the software accompanying the license should probably not be in a public repository.
\item A clearly proprietary license (see Figure~\ref{fig:cases-proprietary}).\footnote{SWHID \swhid{swh:1:cnt:9bffa2d5a63151c8c9bf3d68e9f9445558273612}} Likely, the software that it accompanies should not be in a public repository. However, this case also shows how the variety of licenses found goes beyond FOSS licenses.
\item MIT license written in HTML, with some code for navigation of a webpage.\footnote{SWHID \swhid{swh:1:cnt:c53a6c27009183d8304d26a213b1321bdfc0cb8d}} This document shows how licenses can be in formats other than plain text, and also that when they are in HTML (maybe because they were intended to be in a website) there could be many decorations that make it more difficult to identify them as licenses.
\item ``This is love'' as the only text in the document.\footnote{SWHID \swhid{swh:1:cnt:41a6fc531459dde48d1752f24eae007047361709}} A sample of strange things that can be found in the collection.
\item A \revchange{(now expired)} license token for some software.\footnote{SWHID \swhid{swh:1:cnt:4e5eebfdbebefe990e309ecbdd83842035d3852c}} Another interesting case of weird stuff that can be found in the collection. The filename ``LICENSE'' can perfectly be used for a license token for some software. Arguably the token should not have been published in a public repository.
\item Apparently contradictory licensing information, stating that the software is in the public domain, but also that authors can be contacted to get a license.\footnote{SWHID \swhid{swh:1:cnt:105961e3702324fadaa808457338a984101d6028}}
\item Documentation for a class for handling licenses, in some programming language.\footnote{SWHID \swhid{swh:1:cnt:f3932de6d7f19b26afaa7bc8502c800476c2f0a5}} Even when for a human it is quickly clear the text is not a license, the words and the structure of the text could easily trick heuristics designed to detect licensing information.
\item Document that just says ``free'' (see Figure~\ref{fig:cases-free}).\footnote{SWHID \swhid{swh:1:cnt:fed8329964dd68adcd3dc98dd405950e53614282}} Of course, this is not a license, but still, it could be interesting in the context of other information.
\item License in Chinese, in HTML format.\footnote{SWHID \swhid{swh:1:cnt:60ff9a40c14915b25d265f2bdfb508274b6782fe}} Licenses in languages other than English are perfectly valid, and should be detected when scanning for licensing information. But not being in English and, in this case, not being plain text, makes it really difficult to automatically identify this document as a license.
\item Document that just says ``All rights reserved''.\footnote{SWHID \swhid{swh:1:cnt:ace0bbb7fe0a8677ef5ae001b5da076b2aa666a5}} This can be considered as licensing information, but it has some implications. First, the text is so short that it is difficult to detect except if the tool is looking specifically for it. Second, this would make the software proprietary, and likely should not be in a public repository.
\item Licenses as strings in C++ code.\footnote{SWHID \swhid{swh:1:cnt:9392142a987ee04c3f0d303a58b19df818df86b3}} Yet another example of how the text of a license can come in many different formats.
\item Python code to ``load authors''.\footnote{SWHID \swhid{swh:1:cnt:eb531dc6990ca433ccde3100633780ad55aed22b}} Interesting case that shows unintended files that can be captured with the query heuristic used for composing the collection.
\end{itemize}

 \section{Hands-on dataset usage examples}
\label{sec:usage}

We provide below some hands-on usage examples of the dataset, using the Python language and the Pandas~\cite{mckinney2011pandas} and Scikit--learn~\cite{pedregosa2011:scikit} data science libraries. First, we will show some examples of how the Metadata Files can be used to obtain a characterization of some basic parameters about the Document Collection. Then, we will work with documents in the collection to obtain the word frequency distribution of all of documents in plain text format. Finally, we will use the Annotated Sample to train and test a random forest classifier designed to predict when a document includes a single full-text license. They are just examples of how to work with the dataset, but they do provide valuable insights: the first example shows how most of the characterization in Section~\ref{sec:characterization} was done; the second one illustrates how documents can be processed; and the third one exhibits the kind of analysis for which the Annotated Sample was produced: to be used as ground truth for classifiers and other processors that could work with the whole collection.

\subsection{Working with Metadata Files}

Some information can be extracted from the Metadata Files with simple command-line tools. For example, the number of documents with licensing information according to ScanCode, presented in Section~\ref{subsec:characterization-licenses-found}, was obtained from the \texttt{scancode} CVS file as follows:

\begin{lstlisting}[basicstyle=\ttfamily\small]
cut -d, -f1 blobs-scancode.csv  | uniq | wc
\end{lstlisting}

\noindent
For more complex analyses, scripting in some programming language is usually more convenient. For example, to get a general feeling of the dataset one can look at the \texttt{fileinfo} CVS file in the Metadata Files, and produce some descriptive statistics about it as follows (in Python):

\begin{lstlisting}[style=myPython]
import os
import subprocess
import pandas as pd

stats_csv = f"{dataset_dir}/blobs-fileinfo.csv"
if not os.path.isfile(stats_csv):
    subprocess.run(["unzstd", "--force", stats_csv + ".zst"], \
    check=True)
stats = pd.read_csv(stats_csv)
stats.describe()
\end{lstlisting}
where \texttt{dataset\_dir} is a variable pointing to the local dataset download directory.
Note how we take care of decompressing the relevant CSV file from the dataset distribution.
The results obtained will look like this:

\begin{center}\small
\begin{BVerbatim}
         line_count    word_count          size
count  5.667116e+06  5.667116e+06  6.859190e+06
mean   1.307410e+02  8.610119e+02  1.015964e+04
std    4.511294e+03  1.450112e+04  2.450215e+05
min    1.000000e+00  0.000000e+00  0.000000e+00
255075max    6.373094e+06  7.374871e+06  1.909773e+08
\end{BVerbatim}
\end{center}
They provide a preliminary statistical overview of the different size metrics of all the documents in the dataset (see Section~\ref{sec:findings} for a more refined analysis).

The list of top file types in the dataset, analogous to what we reported in Figure~\ref{fig:documents-filetypes}, can be obtained in tabular form analyzing the same metadata as above, like this:
\begin{lstlisting}[style=myPython]
mime_top = stats["mime_type"].value_counts()\
  .nlargest(20).rename_axis('mime_type').reset_index(name='counts')
mime_top.to_csv(out_data_dir + "/mime_top.csv", index=False)
mime_top
\end{lstlisting}
\begin{center}\small
\begin{BVerbatim}
                    mime_type   counts
0                  text/plain  5721424
1                   text/html   723593
2                  text/x-php    65275
3                 text/x-java    61554
4    application/octet-stream    49195
5                    text/xml    49101
6                   image/png    22912
7            application/json    20327
8        text/x-script.python    15703
9            application/gzip    12367
10                   text/rtf    11956
11                text/x-ruby    11646
12                  text/x-po    11102
13                   text/x-c    11084
14                 text/x-c++    10549
15  application/x-java-applet     6584
16              image/svg+xml     6493
17                 text/x-tex     6286
18     text/x-bytecode.python     4701
19            application/csv     4674
\end{BVerbatim}
\end{center}

\noindent
A similar analysis for character encodings would be:
\begin{lstlisting}[style=myPython]
encoding_top = stats["encoding"].value_counts()\
  .rename_axis('encoding').reset_index(name='counts')
encoding_top.to_csv(out_data_dir + "/encoding_top.csv", index=False)
encoding_top
\end{lstlisting}
\begin{center}\small
\begin{BVerbatim}
       encoding   counts
0      us-ascii  5517915
1         utf-8  1154191
2        binary   121966
3    iso-8859-1    49251
4  unknown-8bit    13597
5      utf-16le     2125
6      utf-16be      143
7        ebcdic        2
\end{BVerbatim}
\end{center}

\subsection{Working with the Document Collection}
\label{subsec:usage-collection}

As an example of how to analyze the actual license documents, as opposed to only associated metadata, we show how to obtain the word frequency distribution of the entire dataset:
\begin{lstlisting}[style=myPython]
def mine_word_frequency(df, out_fname):
    from collections import Counter
    import re
    import string

    WORD_SEP = re.compile(r"\W+")
    word_freqs = Counter()
    text_blobs = df[(df.mime_type == "text/plain") \
      & (df.encoding.isin(["us-ascii", "utf-8", "iso-8859-1"]))]
    for sha1 in text_blobs["sha1"]:
        fname = f"{dataset_dir}/blobs/{sha1[0:2]}/{sha1[2:4]}/{sha1}"
        try:
            with open(fname, encoding="utf-8") as f:
                for line in f:
                    word_freqs.update(
                      (word
                       for word in WORD_SEP.split(line.lower())
                       if word)
                    )
        except ValueError:  # decoding errors
            continue

    with open(out_fname, "w") as csv:
        csv.write("word,frequency\n")
        for (word, freq) in word_freqs.items():
            csv.write(f"{word},{freq}\n")

words_csv = "blobs-wordfreqs.csv"
mine_word_frequency(stats, words_csv)

import nltk
import string
from nltk.corpus import stopwords

nltk.download("stopwords")
stop_words = stopwords.words('english')
stop_words.extend(string.digits)
stop_words.extend(string.ascii_lowercase)

words = pd.read_csv(words_csv)\
  .sort_values(by="frequency", ascending=False)
interesting_words = words[~words["word"].isin(stop_words)]
interesting_words
\end{lstlisting}

\noindent
The code above assumes that \emph{all} license blobs have already been decompressed in the \texttt{blobs/} sub-directory of the dataset download directory.
Doing so is trivial, but note that it will create \DataBlobCountApprox{} files on the filesystem and requires \DataTarExpandedSizeApprox{} of disk space.

The main function is \texttt{mine\_word\_frequency} which will go through all license documents in the dataset to collect the frequency of all words (including stop words).
Note that the approach here is very naive, e.g., with no parallelism involved; better solutions can be found in NLP processing frameworks like Gensim~\cite{srinivasa2018natural}.
At the end of the mining results are serialized in CSV format to \texttt{blobs-wordfreqs.csv}.
That file can in turn be loaded and inspected, after removing English stopwords (according to NLTK~\cite{bird2006nltk}) and one-character ``words'' (although your mileage may vary).
Results were those shown before in Table~\ref{tab:top-words}.

\subsection{Working with the Annotated Sample}
\label{subsec:usage-annotated-sample}

As an example of possible uses of the Annotated Sample, we show below a simple program for producing a random forest model trained to classify documents in two categories: those containing the whole text of a single license and the rest. This code assumes that the \texttt{licen} and \texttt{licens} modules\footnote{ \texttt{licen} and \texttt{licens} are Python modules for dealing with the Document Collection.} are available in the Python \texttt{sys.path}, that the function \texttt{path\_from\_filename}\footnote{\texttt{path\_from\_filename} is a function returning the path of a document in the collection, given its name (SHA1)} is available to the interpreter \revchange{(please note that this is just an example for illustration purposes, not intended for a real analysis, and thus not including optimization techniques that would improve results)}.\footnote{For a full, ready-to-work program, check the file \texttt{truth/random\_forest.py} in the dataset} In this code, \texttt{licenses\_dir} points to the Document Collection and \texttt{truth.csv} to the Annotated Sample.

The program starts by importing all modules needed:

\begin{lstlisting}[style=myPython]
import os
import pandas as pd
from sklearn.feature_extraction.text import CountVectorizer
from sklearn.feature_extraction.text import TfidfTransformer
from sklearn.metrics import (classification_report,
  confusion_matrix, accuracy_score)
from sklearn.model_selection import train_test_split
from sklearn.ensemble import RandomForestClassifier
import licens
import licen
\end{lstlisting}

Then, we load the Annotated Sample into a dataframe, get all documents in it from the Document Collection, and produce a list with strings corresponding to the normalized version of all those documents. Normalization covers: removal of all blanks, lowercasing, and removal of lines containing simple copyright notices:

\begin{lstlisting}[style=myPython]
# Read the Annotated Sample (one line per document) into a dataframe
truth_df = pd.read_csv('truth.csv')
# Compute normalized versions of all license documents in the sample
paths = [path_from_name(name, licenses_dir)
         for name in truth_df['name']]
documents = licens.Licenses(paths,
        cache=None, rebuild_cache = True,
        comps=[licen.CompUnified, licen.CompIdentify])
# Get a list with the normalized text for all plain text documents
text_licen = [licen
              for licen in documents.get_licenses()
              if (licen.utext is not None) \
                and (licen.kind == 'text/plain')]
udocuments = [documents.get_license(licen.filename, licen.name).utext
             for licen in text_licen]
\end{lstlisting}

\noindent
We now produce the source (X) and target (y) values for training and testing the random forest model, and split them in training and testing sets:

\begin{lstlisting}[style=myPython]
# Source values: Vectorized normalized documents (TF-IDF representation)
vectorizer = CountVectorizer(max_features=1500, min_df=5, max_df=0.7)
X = vectorizer.fit_transform(udocuments).toarray()
tfidfconverter = TfidfTransformer()
X = tfidfconverter.fit_transform(X).toarray()
# Target values: True if document includes a singl full-text license
y = [truth_df.loc[truth_df['name'] == doc.name, 'licen'].values[0]
     for doc in text_licen]
# Get training and testing sets
X_train, X_test, y_train, y_test = \
  train_test_split(X, y, test_size=0.2, random_state=0)
\end{lstlisting}

\noindent
Finally, we train and test the random forest model:

\begin{lstlisting}[style=myPython]
# Train a random forest classifier
classifier = RandomForestClassifier(
  n_estimators=1000, random_state=0, n_jobs=-1)
classifier.fit(X_train, y_train)
# Predict which files are full-text licenses for testing sample
y_pred = classifier.predict(X_test)
# Print results of the prediction
print(confusion_matrix(y_test,y_pred))
print(classification_report(y_test,y_pred))
print(accuracy_score(y_test, y_pred))
\end{lstlisting}

\noindent
The result of running this program is shown in Table~\ref{tab:random-forest-results}, obtaining an accuracy of 91.7\%, which is not bad at all for a first try!

\begin{table}
  \caption{Results of running the program that trains and evaluates a Random Model classifier for predicting when a document includes a single full-text license. The accuracy of the model is evaluated as being 0.917.}
  \label{tab:random-forest-results}
  \centering
  \begin{tabular}{l|r|r}
         & \textbf{Negatives} & \textbf{Positives} \\ \hline
    \textbf{True} & 371  & 85 \\
    \textbf{False} & 19 & 782 \\
  \end{tabular}
  \vspace{2ex}

  \begin{tabular}{l|c|c|c|c}
            & \textbf{Precision} & \textbf{Recall} & \textbf{F1-score} & \textbf{Support} \\ \hline
    \textbf{False}    &   0.95  &    0.81  &    0.88  &     456 \\
     \textbf{True}    &   0.90  &    0.98  &    0.94  &     801 \\
  \end{tabular}
\end{table}

 \section{Discussion}
\label{sec:discussion}

In this section we discuss the validity of the dataset for studying licenses (which was the main aim when producing it), the interest of the dataset for practitioners and researchers, the potential for identification of different license texts and variants, and the differences between the different editions of the dataset.

\subsection{Validity of the dataset for studying licenses}

One of the main reasons for compiling the Document Collection was to produce a collection of documents that help to improve the knowledge about licenses for publicly available software. For this, a large fraction of the collection should be composed by licenses, and a large fraction of licenses used for publicly available software should be present in the collection:

\begin{itemize}
\item \textbf{Fraction of the collection composed by licenses}. We have analyzed the contents of the collection from two different perspectives. First, we used the ScanCode tool to analyze the documents in the collection, which gives us an approximate \emph{lower bound} on the number of documents related to licensing that are present in the collection. Second, we manually analyzed the Annotated Sample, to learn what fraction of it was actually composed of licenses.

  The ScanCode analysis presented in Section~\ref{subsec:characterization-licenses-found} showed that most of the documents (70.84\% of the collection) were found to contain at least a license text or notice. Since the information in the \texttt{scancode} CSV metadata file does not detail if ScanCode identified the full text or licenses, or references to them, this number can be interpreted only as ``ScanCode found some licensing information in the document''. Besides, ScanCode heuristics can fail, either by identifying a license or license notice in a document that does not have it, or by ignoring a license or license notice in a document. Thus, the number is only approximate.

  Hence in Section~\ref{subsec:characterization-annotated-sample} we showed that the manual analysis found an error rate of 2.7\% in the identification of full licenses by ScanCode. There are no reasons to think that the rate in license notices is very different. We therefore consider the ratio above (70.84\%) to be a good approximation of the fraction of the collection that includes licensing information.

  The manual analysis performed on the Annotated Sample is consistent with this result. As shown also in Section~\ref{subsec:characterization-annotated-sample}, specifically in Figure~\ref{fig:truth-kinds}, the total fraction of documents found manually to have licensing information is 72.81\% of the sample. Being a random sample, this number can be extended to the whole collection. The manual analysis also showed that 56.87\% of all documents include the text of a single license (as discussed in Section~\ref{subsec:findings-rq1}).

  To summarize, both analyses show that more than two thirds of the collection include documents containing licensing information and that more than half of it contain the text of a single license.
  
\item \textbf{Fraction of known licenses included in the collection}. To estimate this number we consider that the ScanCode LicenseDB collection is a good proxy for the list of known licenses. As we showed in Section~\ref{subsec:characterization-known-licenses}, there are 1879 distinct license texts in this collection, of which we found 1847 (98.29\%) with exactly the same text in the collection. Therefore, we can say that a very large fraction of the most complete list of licenses known to us is present in the Document Collection. It is likely that the few licenses that we could not find with exactly the same text are in fact presence in the collection with similar text variants (e.g., with differences only in blanks or one-line copyright notices). \revchange{However, since the ScanCode LicenseDB has been built incrementally over several decades, with all licenses they have ever encountered, it is also possible that some of these licenses are no longer around, and hence not found in Software Heritage that started archiving in 2015, and initially only GitHub.}
\end{itemize}

\noindent
Based on these considerations, we conclude that the Document Collection includes a very large part of the license texts ever used for publicly available software, and that most of the documents in it include either full texts of licenses or license notices. By extension we argue that the Software Heritage License Dataset presented in this paper is a good dataset to advance the state of knowledge about licenses used for public code.
 
\subsection{Interest of the dataset}

Detecting licenses is an important topic for both industry and academia. It is for industry, because when reusing software components it is important to know about their license terms, and those are codified in the text of the software licenses that come with the component. Given the large quantity of publicly available software components, this task needs automation, and several tools do exist to perform that task (one of them being ScanCode, which we use to automatically annotate the dataset). However, the detection heuristics of these tools, even when carefully tuned over time, still have room for improvement. By exploring our Document Collection, we can understand why: there are many variants of license texts and they are not always captured by those heuristics. Also, there are rare licenses, which are not in the license database used by those tools.

We expect that our dataset can be used to improve this situation. On the one hand, it can be used to improve heuristics for detecting variants of licenses, and to try new approaches for detecting when a file contains a variant of a known license.  On the other, the dataset can be used to find new licenses, not yet considered by those tools. The Metadata Files helps, via the data produced by ScanCode, to understand the large variety of license texts, and the Annotated Sample of the collection shows that  there is still some more variety not captured by state-of-the-art heuristics (because some license files are not properly detected), and that other licenses, unknown to the tools, do exist.

For research and academia, the dataset can be used in different ways. The collection is a good set of documents to work with techniques to identify licenses, to analyze the evolution of licensing information, or to analyze semantic variants of licenses. Since the Document Collection likely includes almost all license texts ever used publicly, it may be an invaluable resource for avoiding the data collection aspect of these studies, and still work with good and comprehensive data about almost any problem related to software license texts.

The Metadata Files allows for even easier study of different aspects of licensing information. It includes detailed information about different aspects of all documents, which may be suitable for studies using machine learning techniques, or statistic analysis. By using it, researchers are spared the automatic annotation process and at the same time use a well known dataset, which should facilitate the comparison with other studies also using it.

Finally, the Annotated Sample provides a detailed, manually curated analysis of thousands of documents, the largest to date and to our knowledge. It can be used as ground truth or training data for studies needing such information. It can also be a starting point for any larger analysis on the whole collection, by first testing specific techniques on the sample, which allows for an evaluation of the technique with reliable data.

\subsection{Identification of license texts and license variants}

To improve the accuracy of tools finding licenses in source code, and to better understand software licensing of publicly available source code, it would be convenient to extract from the Document Collection two specific sub-collections: all documents that include the text of a single license, and a collection with a representative of all semantically different licenses.

\begin{itemize}
\item \textbf{Documents with a single license text}. This collection will be very appropriate to detect licenses in any software. If any file, or part of a file, matched one of these documents, it would certainly be a license file. In addition, given the very large coverage of the Document Collection, this collection would include almost all license texts ever included with publicly available source code.

  Unfortunately building this collection is not trivial. Given the sheer number of documents in the Document Collection, manual analysis would be unfeasible; some kind of automatic classification should be used instead. From this point of view, the very simple approach followed in Section~\ref{subsec:usage-annotated-sample}, training a random forest model to identify documents with a single full-text license, is promising. With no specific fine-tuning we reached an accuracy close to 0.92, which is a very good baseline to build upon. In addition, using other models, such as Large Language Models, that have shown to be very good in similar classification problems of text documents, could significantly improve accuracy. The Annotated Sample could also be extended, to produce more training data, but it is not clear that would improve results significantly.

\item \textbf{Representatives of all different licenses}. In our analysis we have explored how many texts are very similar. Licenses such as MIT or BSD have hundreds of thousands cosmetic variants, different only in blanks, or in one-line copyright notices. Many of them are semantically identical. In this respect, we refer to ``semantically equal'' if the normative text of the license is exactly equal, once you normalize blanks, lower/uppercase, and other aspects such as one-line copyright notices that add nothing to the meaning of the license.

  Having this collection of representatives of these families of semantically equal licenses would allow to focus on the semantic analysis of licenses, and really learn about the variants that introduce meaningful changes. In a software license, even small editions, such as adding ``not'' before a permission may lead to a completely different meaning. Therefore, studying the differences between semantically equal families would help to really understand how varied licenses for publicly available software are.
\end{itemize}

\subsection{Multi-license files}

\revchange{
As shown in Figure~\ref{fig:truth-kinds}, in the Annotated Sample we have found a good number of files with more than one full text license or license notice (over 8\%). This could be expected, because of two main reasons:
}

\begin{itemize}
\item \revchange{Software projects, and in particular FOSS projects, may include pieces of software with different licenses. Licensing is usually done at the component level, and may be done even at the file level. Because reusing software is so common in FOSS communities, in some cases just by copying it into the code base, many software projects end up including software with different licenses. In this case, it is common to inform of this fact in a file with the source code, which includes full text or notices of the several licenses of that source code.}

\item \revchange{Software components can be licensed in full under several licenses (dual or multiple licensing). In this case, the software can be used following the terms of any of those licenses, and this is in many cases informed, again, in a file that includes the full text, or notices, of the licenses that can be chosen.}
\end{itemize}

\noindent
\revchange{
There is a special case of these multi-license files which deserves a specific mention: Debian copyright files. These files, specific to the Debian GNU/Linux distribution (but also present in other popular distributions based on it, such as Ubuntu), intend to include \emph{all} the licenses applicable to a given software package. In the annotated sample we have found about 3.5\% of files which seem to be Debian copyright files (see details in Table~\ref{tab:truth-characteristics}). Even when not all of them are multi-license files, they explain a good fraction of all the multi-license files found.
}

\revchange{
It is also important to mention that, when finding different license texts, we considered only files with a single license, and not multi-license files. This was due to the fact that in multi-license files we cannot tell when the same text for a license is exactly the same, but the combination of licenses and license notices is different, or when we have a really different text. Therefore, it is likely that we are underestimating the number of different license texts in the collection, since some more could be present in multi-license files.
}

\subsection{Differences between dataset editions}

\begin{table}[tbh]
  \caption{Most notable differences between dataset editions}
  \label{tab:dataset-releases}
  \centering
  \begin{tabular}{l|c|c|c}
    Features / Edition
     & \textbf{\DataEdILabel}
     & \textbf{\DataEdIILabel} \cite{msr-2022-foss-licenses}
     & \textbf{\DataEdIIILabel} (current) \\ \hline
    \hline
\textbf{Documents} & \DataBlobCountApproxEdI{} & \DataBlobCountApproxEdII & \DataBlobCountApproxEdIII \\ \hline
\textbf{Filenames}           & \OK & \OK          & \OK \\ \hline
    \textbf{Fileinfo}            & \KO & \OK          & \OK \\ \hline
    \textbf{ScanCode (summary)}  & \KO & \OK          & \OK \\ \hline
    \textbf{ScanCode (full)}     & \KO & \KO          & \OK \\ \hline
    \textbf{Origin (sample)}     & \KO & (incomplete) & \OK \\ \hline
    \textbf{Origin (count)}      & \KO & \KO          & \OK \\ \hline
    \textbf{Earliest commit}     & \KO & (incomplete) & \OK \\ \hline
    \textbf{Manual annotation}   & \KO & \KO          & (sample) \\ \end{tabular}
\end{table}

As mentioned in the Introduction, this dataset has been published (at the time of writing) in three editions over time, labeled chronologically respectively \DataEdILabel, \DataEdIILabel, and \DataEdIIILabel{} (the version described in this paper).
Table~\ref{tab:dataset-releases} highlights the most notable differences between the various dataset releases up to the one described in this paper.

Focusing on the differences between the current (\DataEdIIILabel) and previous (\DataEdIILabel) editions we observe that the corpus size \revchange{has grown by $\approx\,$6\%, with the addition of almost 0.5 million new documents}, corresponding to one full year of additional source code crawling by \SWH. The filename regular expression used for selecting license documents has not changed between these two editions, so organic growth of the \SWH archive is the primary reason for corpus growth.
In addition to organic growth of data sources that were already crawled by \SWH at the time of the previous dataset release, new archive data sources (forges, package repositories, etc.) have also been added; details are documented on the ``archive changelog'' page.\footnote{\SWH archive changelog page: \\
  \url{https://docs.softwareheritage.org/devel/archive-changelog.html} (accessed 2022-11-10)}

Other significant differences between the two most recent editions are:
\begin{itemize}

\item License popularity information has been extended and now also includes the number of software origins (e.g., VCS repositories) in which each license document in the dataset has been observed in.
  In the previous version only the number of \emph{commits} in which a licensed document has was available.

\item The dataset now includes full ScanCode license and copyright scanning results in JSON format.
  This complements the few summary fields (license and score) that were included in previous versions and are still available in this version (as they are easier to load into tabular processing engines).

\item Metadata inconsistencies about the provenance of license documents have been resolved.
  Earliest commit information is now available for all license documents in the dataset.
  Origin information (both for origin samples and for the number of origins) is missing for just \DataBlobNoOriginPct\% of the license documents in the dataset (down from 10\% in the previous version), making it a quantitatively negligible issue (see Section~\ref{sec:threats} for a discussion of this potential threat to validity).

\item In addition to automated annotation of the dataset based on various tools (e.g., ScanCode) the dataset now includes manual annotation for the Annotated Sample (\DocumentsManualCount{} documents), as discussed extensively in Section~\ref{subsec:usage-annotated-sample}.

\end{itemize}

 \section{Threats to validity}
\label{sec:threats}

\subsection{Internal validity}

Datasets built from large amounts of real-world data tend to be noisy and
contain bogus data (non-license files, in this case) amid legitimate data.
In this dataset, rather than thoroughly
trying to clean up license documents incurring the risk of false negatives, we have
decided to \emph{augment} them with extra metadata that enable researchers to
filter data downstream.
We have already observed in Section~\ref{sec:methodology} how
to restrict the filename pattern if desired. Similarly, researchers can
filter on MIME types (e.g., if only interested in textual files) or on length
metrics (e.g., only keep oneliner files to focus on copyright notices or
machine-readable SPDX tags). Study-specific filtering is also best left to
dataset users; the dataset provides several types of metadata to implement it
in practice.

A minor inconsistency in the dataset comes from the incompleteness of
origin-related metadata (sample origins and number of origins), which are missing for a tiny fraction of documents in the dataset (\DataBlobNoOriginCount{} blobs, or \DataBlobNoOriginPct\% of the full corpus).
The amount is so marginal that we do not consider it to be a significant threat to validity.

Also, due to the ease of
forging Git timestamps~\cite{flint2021timepitfalls, swh-provenance-emse}, some earliest commit metadata are bogus having
timestamps set to the UNIX epoch. Both metadata coverage (which remains very
high) and timestamp quality can be improved by cross-referencing license blobs
to external data sources thanks to the persistent identifiers used in the
dataset as keys.

\subsection{Construct validity}

There is no guarantee that all license blobs in the dataset contain license
texts considered open/free by OSI/FSF, only that they come from public code. If
relying on ScanCode as ground truth is acceptable, \texttt{scancode} metadata
in the dataset can be used for filtering on OSI/FSF approved licenses.
To do so, the machine-readable SPDX license list\footnote{\url{https://spdx.org/licenses/}, accessed 2022-11-10} can associate the SPDX values in the dataset to identify those licenses considered Free Software by the FSF and which are OSI-approved.
Otherwise the free software/open source determinations
will need to be done independently on their own by dataset users.

\revchange{We have considered filenames likely to include licensing information, and as much as possible, only licensing information. Therefore, we have omitted files that in many cases \emph{also} include licensing information, such as \texttt{README} or \texttt{README.md} files, but in many others do not, and usually include other text not related to licensing. We have also omitted source code files, which in many cases include a license text or notice as a comment in the file header. This means that the dataset does not include all files in \SWH with licensing information, but only those that we considered more likely to have licensing information.}

\revchange{Therefore, license texts and license notices that only appear within source files or in files with filenames not captured by our heuristics are very likely to be underrepresented in the dataset.
This applies to, for example, both the recommended GPL notice {\em ``This program is free
software [\ldots] under the terms of the GNU General Public License [\ldots]''}
and SPDX tags~\cite{stewart2010spdxspec} like ``SPDX-License-Identifier:
GPL-3.0-or-later'' when they are included only as comments at the beginning of
source code files. We considered that including those files would significantly increase noise in the dataset, by including files with no licensing information. Besides, reliably identifying licenses in those cases would be more difficult, since they are mixed with other text of very varied nature.}

Thoroughly extracting license- and copyright-related information from all files archived by \SWH and including them in the dataset is left as future work.

\subsection{External validity}

By its own nature, the dataset provides an incomplete snapshot of reality; as
such we do not claim full generality nor representativeness of all existing license
variants. The reality is a moving target, with new license variants constantly
released as part of public code. The archive we started from is not full-encompassing
either. Still, and to the best of our knowledge, this is to date the largest
publicly available dataset of (public code) license variants. We plan to
mitigate this risk by periodically making new dataset releases available, as we
have done up to now and once again with the novel dataset release documented in this paper.

 \section{Related work}
\label{sec:related}

\revchange{
There are many studies, datasets and software related to the identification of licensing information in source code. In this section we will review some of the most relevant.
}

\subsection{Analysis on source code licensing}

\revchange{
  The analysis of licensing information in source code was an already an active area of research about 15 years ago. In 2009 Germán et al. identified different practices of code reuse under the GPL license, by analyzing the license of packages and how they combined different licenses with the GNU GPL~\cite{german2009:reuse-gpl}. That same year Germán et al. presented a more ambitious analysis, also by checking licenses in source code packages, of the more general problem of licensing mismatches, identifying several integration patterns depending on the licenses used~\cite{german2009:license-mismatches}. Another two seminal papers on FOSS licensing have been authored by Germán et al.~\cite{german2010:licensing-distributions} and di Penta et al.\cite{penta2010:evolution-licensing}. The first one is focused on the problem of license compatibility, in this case in the context of software compilations. It presents a detailed analysis of all source code files in the 1475 source packages of a Linux distribution, using the Ninka tool (see some details about this tool below), detecting some potential license incompatibility problems, and showing the diversity of licenses in a real distribution.  The second one analyzes in detail, using also  Ninka, the evolution of licensing information the all source code files of 6 popular FOSS components, uncovering how their licensing terms went through several important changes as those components evolved.}

\revchange{Other later studies relevant for the dataset presented in this paper can be cited. For example, Manabe et al. used license inclusion graphs for analyzing how different licenses are used together in the same software package~\cite{manabe2014:license-packages}. Maryka et al. offered the first large-scale study on license variability, studying different BSD and MIT licenses variants, and when those variations have legal meaning or are just different ways of writing the same text~\cite{maryka2015:maryka}. In some sense, this is a study that could be replicated on the much larger collection of license variants present in our dataset. Vendome et al. explored the problem of how and when developers decide to change software licenses, also based on an actual analysis of licenses in source code~\cite{vendome2015:developers-change-licenses}. In a later study, Vendome at al. performed a license analysis of source code hosted in GitHub, extending by some orders of magnitude the sample size of licensing source code analysis~\cite{vendome2017githublicenses}. Another relevant precedent is, to our knowledge, the first study to use machine-learning techniques to find and identify changes to software licenses with legal meaning, such as exceptions to the license conditions~\cite{vendome2017licexceptions}. For that, the authors use a large sample of license texts found in source code.
}

The Debsources dataset \revchange{(discussed below in the related work section dedicated to datasets)} has been used to conduct a large-scale, longitudinal study of license evolution over multiple decades of time in the Debian distribution~\cite{debsources-ese-2016}, which would be interesting to replicate using our dataset, to compare and contrast a system/infrastructure ecosystem like Debian with a much broader view of public code.

\revchange{
It is worth mentioning that most, if not all, of these cases focus on the analysis of source code found in the headers of source code files, making them somewhat different to the technique used for the collection of our dataset, based on the identification of files that usually include only license information.
}

\subsection{Source code datasets}

The Software Heritage (SWH) graph dataset~\cite{swh-msr2019-dataset}, which we
used to select license blobs, is a large dataset underpinning the SWH archive.
It stores information analogous to those captured by version control systems
(VCS), minus actual file contents. It can be used in conjunction with the
license dataset presented here, joining information via SWH identifiers (the SWHIDs present in the main index and various metadata tables of this dataset).

World of Code (WoC)~\cite{ma2021woc} is a large dataset and analysis infrastructure,
available on demand to researchers to conduct large-scale mining experiments on public code. WoC is larger (although not a strict superset) than our initial
data source and can be used in conjunction with this dataset to find additional
origins/occurrences of licenses blobs of interest.
Due to its focus on license-related information, our dataset is smaller than WoC, and can
be easily self-hosted, and comes with several relevant metadata precomputed (e.g.,
ScanCode results, manual annotations).

Boa~\cite{dyer2015boa} is an infrastructure and a set of accompanying datasets for analyzing source code artifacts coming from public code at a large-scale.
Some of the datasets hosted at Boa contain source code files and commit information coming from publicly-available VCSs, and in particular those hosted on GitHub.
Boa indexes the text of source code files and, as such, can be used to conduct license-related analysis.
Due to its focus on indexing abstract syntax trees (ASTs) Boa would be best suited to extract license/copyright-related information from source code file headers, which are missing from this dataset.
However, the largest VCS repository dataset available on Boa is smaller and has a much smaller coverage than this dataset (i.e., Boa was last updated in 2019 with 8 million repositories vs.~the \DataSwhOriginCountApproxM{} millions in 2022 in our dataset).

GHTorrent~\cite{GHTorrent} is a dataset of archived GitHub REST API events.  It
contains information about public GitHub projects, but as of today does not
include the license that GitHub detected as the main license of a given
project. Being source code out of the scope of GHTorrent, license texts cannot be found
in it neither.

\subsection{License detection tools}

\revchange{
  Detecting licenses and provenance information in software packages, and matching them to how they can be used alone or in combination with other packages, is a field of great industrial interest, but difficult and complex to deal with. During the last years the tools and systems developed to fix this industrial need have evolved quickly~\cite{ombredanne2020sca}.
}

\revchange{
  The ScanCode toolkit~\cite{scancode-toolkit, ombredanne2020sca} is a set of tools to help in the scanning of source code to determine its provenance and licensing terms. It uses three different approaches to find licensing information: pattern matching with small handcrafted text patterns; probabilistic text matching, using some similarity metric to find the closest matching license text or notice; and exhaustive pairwise comparisons to find similar licenses (diff technique). This makes it very successful at finding not only known licenses, but also other licenses similar to them.
}

\revchange{
  There are other tools for identifying licenses in source code, such as FOSSology~\cite{gobeille2008fossology} (which uses a variant of the first approach, an algorithm called the Binary Symbolic Alignment Matrix), ninka~\cite{german2010ninka} (also based on the first approach, using regular expressions), and GitHub Licensee~\cite{github-licensee} (which uses the second approach). Even if we are not aware of third-party scientific benchmarks comparing these and other tools, it is generally assumed in the industry that ScanCode is the state of the art in identifying licensing information. This was fundamental in selecting it for analyzing our collection with it. A more complete list of tools for license detection is maintained by the Debian project.\footnote{Debian Copyright Review Tools: \url{https://wiki.debian.org/CopyrightReviewTools}}
}

\revchange{
Other related tools are devoted to other use cases related to licensing, such as detecting conflicts and license incompatibility. For example OSLDetector~\cite{zhang2021:osldetector}, a library detector capable of selecting library versions with a certain set of features, also checks license conflicts. LiDetector~\cite{xu2023:lidetector} uses a learning-based method to automatically identify meaningful license terms from licenses, employing probabilistic context-free grammar to infer rights and obligations for incompatibility detection. 
}

\revchange{
  Systems that collect and store licensing information for large collections of FOSS packages are also emerging during the last years. Some prominent examples of this trend are Libraries.io~\cite{libraries.io}, which provides information about millions of FOSS packages, including licensing information, the CodeMeta Project~\cite{codemetaproject}, focused on producing code metadata, including licensing information, and ClearlyDefined~\cite{clearlydefined}, which provides a mechanism for harvesting and curating licensing data about FOSS packages using ScanCode, FOSSology, and other tools.
}

\subsection{Datasets with license texts}

The ScanCode LicenseDB~\cite{scancode-licensedb} is a public database curated by the
ScanCode authors that lists all the licenses they have encountered in the wild
during the constant tuning of the ScanCode detection heuristics. As of March 2022 it includes 1879
different \emph{canonical} license texts which are used as comparison
reference, but does not provide all variants of them as we do with this
dataset; nor it provides associated metadata.

Both the Open Source Initiative (OSI)
and the SPDX project maintain analogous public databases~\cite{osi-licensedb,
  SPDXLicences} covering the canonical texts of, respectively, OSI-approved and
SPDX-named licenses, for about 500 texts in total.
Both the OSI and SPDX databases contain only \emph{canonical} license texts, rather than all license variants encountered in the wild as our dataset does.

The Debsources dataset~\cite{debsources-ese-2016} contains the complete source code as well as metadata about all packages shipped as part of the Debian distribution, both at present time and historically over time.
Metadata in the dataset were extracted automatically using several tools.
The used tools include license-detection tools, and in particular ninka and FOSSology (but not ScanCode, which we used for this dataset).
The Debsources dataset comes from a curated software distribution (Debian) and is hence likely to contain more high-quality data, at the price of a much smaller coverage.
For comparison, our dataset contains license documents coming from \DataSwhOriginCountApproxM{} million software origins, whereas the Debian distribution in all its releases has shipped a bit less than 50 thousand source packages.

\smallskip

In summary, this dataset appears to be unique in nature and size, filling an
unattended niche of license-related data about public code.
It can also be used in synergy with preexisting datasets
about FOSS and public code.

 \section{Conclusion}
\label{sec:conclusion}

We have introduced a large-scale dataset of open source license texts. It consists of a collection of \DataBlobCountApproxM{} million unique documents archived from public code and carrying filenames related to software licensing terms; of metadata files to simplify some kinds of analysis; and of a manually annotated random sample suitable  for validation and training purposes. We have described it in detail both to help in its replication, and to enable its use for further research by any third party. We have shown how the dataset is useful for its intended usage, including almost all known license texts, and being composed fundamentally of documents with licensing information. We have also explained why the collection has such a variety of licenses, and illustrated how the different components in the dataset can be used.

\paragraph{Future extensions}

As future work we intend, on the one hand, to keep the dataset current with the
constant evolution of archived public code, gathering license texts from
additional data sources.  On the other hand we will explore adding to the
metadata precomputed text representations of the entire corpus that are
commonly needed for natural language processing (NLP) and machine learning
analyses, such as word embeddings, latent semantic indexes, and other vectorial
text representations.

 \section*{Conflict of interest}

The authors declared that they have no conflict of interest.

 \section*{Acknowledgements}

This work was made possible by Software Heritage, the great library of source code: \url{https://www.softwareheritage.org}.

The authors would like to thank Valentin Lorentz from the Software Heritage engineering team for his help in releasing the new version of the license dataset documented in this paper and streamlining the dataset publication process.

\end{document}